\documentstyle[12pt,aaspp4]{article}
\begin{document}

\title{Chandra and XMM Observations of the ADC Source 0921-630}

\author{T. R. Kallman\altaffilmark{1}, L. Angelini\altaffilmark{1,2}, 
B. Boroson\altaffilmark{3}  and J. Cottam\altaffilmark{1}}

\altaffiltext{1}{NASA Goddard Space Flight Center, LHEA, Code 665, Greenbelt, MD 20771}
\altaffiltext{2}{University Space Research Association}
\altaffiltext{3}{Harvard-Smithsonian Center for Astrophysics, 60 Garden St., 
Cambridge MA 02138}

\begin{abstract}
We analyze observations of the low mass X-ray binary 2S0921-63 obtained with 
the gratings and CCDs on Chandra and XMM.  This object is 
a high inclination system showing evidence for an accretion disk corona (ADC).
Such a corona has the potential to constrain the properties of the heated accretion
disk in this system, and other LMXBs by extension.  We find evidence for 
line emission which is generally consistent with that found by previous 
experiments, although we are able to detect more lines.  For the first time 
in this source, we find that the 
iron K line has multiple components.  We set limits on the line widths and  
velocity offsets, and we fit the spectra to photoionization models and 
discuss the implications for accretion disk corona models.  For the first 
time in any ADC source we use these 
fits, together with density constraints based on the O VII line ratio, 
in order to constrain the flux in the medium-ionization region of the 
ADC.  Under various assumptions about the source luminosity this 
constrains the location of the emitting region.  These estimates,
together with estimates for the emission measure, favor a scenario in which 
the intrinsic  luminosity of the source is comparable to what we observe.

\end{abstract}

\section{Introduction}

In spite of much study, understanding of the geometry of the gas flows and
circumstellar gas in low mass X-ray binaries, (LMXBs) remains uncertain.
The conventional picture consisting of a neutron star, low mass companion, 
and accretion disk is heavily influenced by analogy with cataclysmic variables.
There are fewer direct observational constraints on the accretion disks in LMXBs
than in other accretion driven systems.
A major contributor to this uncertainty is the influence of X-ray heating, 
which makes circumstellar gas a mirror for X-rays rather than a 
medium which can be studied via its intrinsic emission.
In addition, dilution of the circumstellar emission 
by the strong continuum emanating from the neutron star adds to 
the difficulty of finding signatures of the disk or other structures in 
the spectrum.  

Useful clues to the geometry of LMXBs may be provided by the sources 
in which we are likely observing close to the plane of the accretion disk and 
binary orbit.   These sources may play a role in our understanding 
of LMXBs which is analogous to the role played by Seyfert 2 galaxies in the 
understanding of active galaxies; the configuration of the circumstellar 
gas can be studied when the direct X-rays from the continuum 
source are at least partially blocked.   
In such high inclination systems the presence of partial eclipses by the 
secondary star suggest the presence of an extended component of 
X-ray emission, likely due to scattering and emission from a cloud of 
highly ionized gas.  The X-rays from the neutron star may provide sufficient
heating and ionization to account for the properties of this 
accretion disk corona, or ADC (Begelman, Mckee and Shields 1982; 
Begelman and McKee 1983).  In addition, 
the heated gas may flow outward in a wind (Woods et al., 1996; 
Proga and Kallman 2002) which can potentially affect the mass budget 
of the system. The existence of accretion disk coronae has been substantially confirmed,
and the column density and size have been constrained, by fitting 
of models to observed X-ray light curves.
Simple models treat the corona as a 
spherical cloud (White and Holt, 1982; McClintock, et al., 1982;
Mason, 1986) in which the X-rays are reflected into our line 
of sight by scattering alone.  These models have also constrained
the size and shape of the outer rim of the accretion disk, although the 
results depend on the assumed geometry for the ADC.

In addition to scattering continuum photons, the corona should also radiate 
in atomic emission features arising from partially ionized material.
Such emission has been predicted by, e.g. Kallman and White (1989), and 
by Vrtilek, Soker and Raymond (1994), Ko and Kallman (1994).  
Strong line emission is expected under a variety of assumptions, and that 
the line emission (as measured by equivalent width) is strongest when the 
inclination of the binary system is closest to 90$^o$.  Observations
of line emission have the potential to provide a sensitive test
of the hypothesis that the corona is heated and ionized entirely by 
X-rays from the central neutron star, since the ionization balance and 
emission measure together serve to constrain both the gas density and 
the size of the emission region.

In order to carry out such a test we have used the gratings and CCD detectors 
on both the Chandra and the XMM satellites to observe the eclipse spectrum of
2S0921-63, a well known LMXB which shows the approximately sinusoidal 
lightcurve indicative of an accretion disk corona.  
Its period of 9.01 days makes it one of the longest 
period LMXB systems, and also makes it particularly suitable for 
eclipse studies owing to the duration of the eclipse.
EXOSAT observations of the spectrum through an orbital cycle 
(Mason et al. 1987) showed that the spectrum softens 
during eclipse.  The spectrum was fit to a hard power law, 
with a photon index of 1.16, and a column density of 1.4$\times 10^{21}$
(although Mason et al. find evidence for contamination at low energies 
which makes this value somewhat uncertain).
The eclipse ($\leq 50 \%$ of the uneclipsed flux) lasted
approximately 80 ksec.
The relatively low X-ray to optical flux ratio 
and many recorded optical dips  (e.g. Branduardi-Raymont et al., 1981, 1983; 
Chevalier and Ilovaisky, 1981, 1982) are 
further evidence for a  hidden X-ray source.
Cowley et al. (1982)  estimate the distance to be $\simeq$7 kpc,
implying a luminosity of 2.4 $\times 10^{35}$ erg s$^{-1}$.

2S0921-63 was observed with ASCA on several occasions, 
including a total 
of 27 ksec of good data obtained during eclipse.   Fits to these 
data reveal a hard power law 
continuum and column density comparable to those observed by EXOSAT.  
This spectrum is significantly harder than the spectra typically observed 
from low inclination LMXB  systems.  The spectrum contains 
an iron K line at an energy of 6.75 keV, intermediate between the He-like 
and H-like ionization states of this element.  This line is unique among 
LMXBs for its equivalent width; both ASCA and EXOSAT (Asai, 2000; Gottwald et al.,
1996) measure $\simeq 120$eV for this quantity, while typical 
LMXBs show much weaker lines, $\sim 40-80$ eV. The flux in the 2-10 keV band 
in the ASCA spectrum of 2S0921-63 is 3.2 $\times 10^{-11}$ erg cm$^{-2}$ s$^{-1}$.
In addition, the ASCA spectrum contains evidence for  line emission in the 
0.5-2 keV band, not previously published.

In the following sections we describe our observations of this source using 
Chandra and XMM, the extraction and analysis 
procedures, results of model fits, and their implications for  
understanding of 2S0921-63 and for ADC sources in general.

\section{Observations}

\subsection{Chandra}

2S0921-630 was observed by Chandra using the HETG and ACIS-S detector between 
UT 15:23:49 2001-08-02 and UT 13:01:13 2001-08-03. The observation lasted 
approximately 74000  seconds.  In order to best 
eliminate the influence of any direct X-rays from the neutron star,
the observation was centered on eclipse phases 0.97 -- 1.07,
based on the ephemeris: 
$\phi={\rm JD}2446249.18^+_-0.01 + n \times 9.0115^+_-0.0005$
(e.g. Mason et al., 1987).  Although the eclipse light curve is 
gradual, an observation of this duration fits  within the 
time when the flux is less than $\simeq 60 \%$ of the uneclipsed value.

Figure 1 shows the HEG and MEG light curves during our observation, plotted against 
orbital phase using the ephemeris of Mason et al (1987).  The data have been 
binned into 100 second bins.  A secular
variability is apparent, resulting in a factor $\simeq$2 increase in the counting 
rate between the beginning and the end of the observation.  This variability is more
pronounced in the MEG than in the HEG, suggesting that the orbital modulation is 
stronger at lower energies.  The statistical 
quality of the HETG spectrum is not sufficient to allow sensitive searches for spectral 
variability associated with this increase.
The mean first order counting rates were .19 counts/s in the HEG
and 0.33 counts/s in the MEG.   Data was reduced using the standard analysis 
pipeline and Ciao. We add the + and - first order spectra together and analyze 
only these orders.  The zero order ACIS-S spectrum showed clear evidence 
of pileup and will not be discussed further in this paper.  
The spectra from the HEG and MEG were fitted 
simultaneously, using the xspec package (Arnaud 1996). 
Best fit values are found by minimizing the value of the C statistic.  Here and 
in the fitting of the XMM data we quote the values of best fit parameters 
together with the errors derived from stepping the parameter in question and
determining the $\Delta$ C statistic or $\Delta\chi^2$=10 value.  All model 
fits attempt to account for the  ACIS low energy quantum efficiency contamination
by using the acisabs model in the post-release version of xspec 11.2.

\subsection{XMM-Newton}

2S0921-630 was observed by XMM-Newton between UT 11:39:00  2000-12-20 and UT 03:24:49 2000-12-21. 
Data were taken simultaneously for the EPIC detectors MOS
and PN (using the medium filter) and with the RGS.  All data were 
processed with the XMM-Newton Science Analysis Software, 
SAS 5.3, with the corresponding calibration files available for that version.  
The RGS covers a wavelength range of approximately 6 to 38 $\AA$ 
(E=0.35 -- 2.5 keV) with a resolution of 0.05 $\AA$ 
and a peak effective area of about 140 cm$^2$ at 15 $\AA$ (0.826 keV).  
The observation was separated into two intervals, one lasting 
7334 seconds, and a second lasting 61878 seconds.  Here we discuss only the 
data from the second interval.  Figure 2 shows the light curve of the 
PN counts, plotted against 
orbital phase using the ephemeris of Mason et al (1987).  The shape of this  
curve corresponds crudely to the shape of the Chandra HETG lightcurve in the 
local maximum near phase 0.965, minimum near 1.01, and overall secular increase.
The mean counting rate was .25 counts/s in RGS 1 and 0.32 counts/s in RGS 2.  
The first order spectra from the RGS 1 and 2 are fitted 
simultaneously, using the xspec package (Arnaud 1996).

\section{Results}

\subsection{ASCA}

The results from fitting to the low energy portion of the ASCA spectrum 
have not been previously published, and so we present them here prior to the presentation 
of the other results.
The results of fitting the 27 ksec of data contained during eclipse 
on 04-05-1994 06:09:12 are shown in Table 1.  The spectrum was fit to a cutoff power law 
plus narrow Gaussian emission features.  The energies of the features are consistent 
with those expected from the L$\alpha$  and L$\beta$ lines of Ne X, and the 
L$\alpha$ lines of  Si XIV and S XVI.  
The O VIII L$\alpha$ line is 
not statistically significant in this spectrum.   In addition, the fit requires 
a (single) iron K line at 1.82 $\AA$ (6.8 keV), corresponding approximately to the He-like 
ionization state of this element.  The measured width of this line is 0.5 keV (Asai et al. 2000), 
which is likely due to the effects of blending.  These results foreshadow those found by 
the other instruments described in this paper.

\subsection{Chandra HETG}

In order to better understand the contributions of various components to
the spectrum we fit the data from all the various instruments with 
several trial model spectra, with progressing physical sophistication.
The simplest model consists of a single power law plus absorption by 
neutral material.  The results of such a fit to the Chandra HETG data are 
shown in Table 2. Figure 3 shows the ratio of the data to the model for such
a fit.   The HEG and MEG are plotted separately.  
The spectra have been binned for the purpose of plotting only.
The best fit power law index ($\gamma$) is  1.01$^{+0.15}_{-0.9}$, which is consistent with that
found by ASCA.  The neutral column is 1.6 $\times 10^{21}$ cm$^{-2}$, 
which is similar to that found by EXOSAT averaged over orbital phase.  This 
value is affected by the treatment of the acis contamination in xspec; 
when the acisabs model is omitted from the analysis the inferred column increases 
to 2.05 $\times 10^{21}$ cm$^{-2}$, other parameters are unchanged. 

Figure 3 shows that, although the fit to the continuum is adequate in the 
Chandra 1-15 $\AA$ (0.8 - 7 keV)  region, there are significant residuals 
near 12, 10.6, 8.4, 7.1, 6.2, 4.7, and 1.7-1.9 $\AA$ (1, 1.3, 1.44, 1.72, 2, 2.6, 6.4 -- 7 keV), 
which correspond to emission lines due to Ne X L$\alpha$, Ne X L$\beta$, Mg XII L$\alpha$ and L$\beta$,
Si XIV L$\alpha$, S XVI L$\alpha$, and Fe I-XXVI, respectively.  
Not all of these lines are statistically significant; Table 2  
shows the effect of adding multiple narrow Gaussian lines to the absorbed 
power law model at the energies of the significant lines.   
This results in a reduction of the C statistic  from 9390 to 9009 for 8384 PHA bins.
All the line fits are consistent with narrow lines ($\sigma \leq 0.005$ for the 
stronger lines), corresponding to Doppler widths less than 1500 km/s.
The flux in the 2-10 keV band 
in this spectrum is 3.8 $\times 10^{-11}$ erg cm$^{-2}$ s$^{-1}$.
Also given in table 2 and all later tables are the line IDs, and the log of the 
probability of random occurrence according to an F test.

The HETG spectrum can also be fitted by a detailed atomic simulation 
of a photoionized gas.   This was done using models calculated with the 
XSTAR code (see, e.g. Kallman and Bautista, 2000; Kallman et al., 1996), which assumes 
that the gas is optically thin, and that the ionization is solely due to 
continuum X-rays from the neutron star.  Such fits constrain the 
ionization parameter and normalization to be log($\xi$)$\simeq$ 4.5  and $\simeq$300, 
respectively.  Figure 3 and Table 3 show the 
best-fit model.  The statistical quality of the fit is 
significantly better than for the Gaussian fits, C statistic=8731 for 8384 PHA bins.
The statistical quality of the data is not sufficient to 
tightly constrain the abundances of the emitting ions; formally 
we obtain 0.1, 0.28, 0.30, 0.33, 0.16, 0.25, 0.22, 2.3e-3, 
for O, Ne, Mg, Si, S, Ar, Ca, Fe, respectively, relative to solar (Grevesse 
et al., 1996).

\subsection{XMM}

The XMM data includes the two grating datasets, two EPIC-MOS, and the EPIC-PN.
Of these, the RGSs provide high resolution 
($\varepsilon/\Delta\varepsilon \sim$ 300) over the 0.5 -- 2 keV energy band, 
while the EPIC detectors provide moderate resolution 
($\varepsilon/\Delta\varepsilon \sim$ 20 -- 50) over the 0.5 -- 10 keV energy 
band.  Owing to pileup of the Chandra zero order, the EPIC spectra provide our only
insight to the broad-band spectrum of 2S0921-63.

\subsubsection{EPIC-MOS}

The MOS provides moderate energy resolution and sensitivity over the 
entire XMM energy band.  Figure 5 shows a fit to the 
global spectrum obtained using a single power law with neutral 
absorption, both the measured and modeled counts and the ratio of 
data to model. Although we have fitted to both MOS detectors simultaneously, 
only MOS 1 is shown in this figure for clarity.  
This shows that the global shape of the spectrum is adequately 
reproduced, and that there are residuals near 6.5-7 keV corresponding to the
iron K lines, in addition to smaller residuals at lower energies, 0.5 -- 2 keV,
corresponding to the K lines from medium-Z elements.  
In  Table 4 we show the parameters of the power law only fit, and for a fit 
to a power law plus Gaussians at the energies of the iron K lines.  We defer 
a discussion of the K lines from medium Z elements to the section on 
RGS fitting.  Table 4 shows that the energies of the iron lines are 
consistent with those of 
the He-like line near 1.82 $\AA$ (6.8 keV), and the H-like line at 1.78 $\AA$ (6.97 keV).  
The neutral-like fluorescence line (Fe I-XVII) at 1.94 $\AA$ (6.4 keV) appears in the 
residuals but is not statistically significant in this fit.
The ratios of the strengths of the lines, and their energies, are 
consistent with those found by the Chandra HETG.  The flux in the 2-10 keV band 
in this spectrum is 5.9 $\times 10^{-11}$ erg cm$^{-2}$ s$^{-1}$.

\subsubsection{EPIC-PN}

The PN provides superior energy resolution and sensitivity at higher 
energies, but owing to calibration uncertainties we can find no satisfactory fit to the 
global spectrum is obtained using any of the spectral shapes that we 
use for any of the other detectors.  In order to avoid this problem, and 
still make use of the high energy part of the spectrum, we fit the PN 
spectrum only at energies between 5 and 10 keV.  In this energy 
band we are able to obtain acceptable fits.  The counting rate is 15.2 counts s$^{-1}$
in this energy band.  The upper panel of figure 6 shows the ratio 
of the PN spectrum to a simple model consisting of a power law 
with index 1.2.  This shows residuals 
at the energies of the 3 iron lines found by the MOS and the Chandra HETG, 
but in addition there is a monotonic decrease in flux above 7 keV, and 
emission features at energies above 7.1 keV.  This spectrum appears superficially 
very similar to the ASCA spectrum of 4U1822-30 in this energy range (White, 1996).
This spectrum does not show the 
behavior expected from Compton reflection, which is a flattening  of the 
power law at high energies.  Instead, the spectrum is apparently steepening between 
7 and 10 keV and contains narrow features which are statistically significant.
 We obtain an acceptable fit this spectrum to 
a power law with index 1.2 and a cutoff at 7 keV, Gaussians at the energies 
of the iron lines: 1.94, 1.85, and 1.78 $\AA$ (6.4, 6.7, and 6.97 keV), plus 
additional Gaussian emission lines at 1.59 and 1.51 $\AA$ (7.8 and 8.2 keV).
This model fit gives  $\chi^2/\nu$=874/1028.
We have also tested this model against the MOS data in this energy range 
and found it to be acceptable.
We have tried fitting this portion of the spectrum to a single power law plus 
atomic absorption as might be expected from an iron K edge or ensemble of edges.
The best fit to a single power law continuum plus Gaussian emission at the energies of the 
iron K$\alpha$ lines  has absorption edges at 7.1 and 8.4 keV.
This results in a fit which has $\chi^2/\nu$=902/1032, which is significantly greater than 
the fit to a curved continuum plus Gaussian emission.
The reason for this is the presence of narrow features at 1.59 and 1.51 $\AA$ (7.8 and 8.2 keV).

The 1.59 $\AA$ (7.8 keV) line wavelength corresponds approximately to that of 
the 1-3 resonance line of Fe XXV, 
or to the 1-2 resonance  line of He-like Ni. The 1.51 $\AA$ (8.2 keV) line is consistent with 
the 1-4 resonance line of Fe XXV, although its strength relative to that of the 
1.85 $\AA$ (6.7 keV) 1-2 line is greater than expected if this is the correct identification.
We can clearly rule out the K$\alpha$ line of Cu, at 1.54 $\AA$ (8.05 keV), which is an 
instrumental feature known to appear in some  PN datasets.
The results are shown in the lower panel of figure 6 and Table 5.
The PN data allows a constraint of the widths of the iron K lines; for the 
1.94 $\AA$ (6.4 keV) line we can marginally exclude  velocities less than .15 keV, corresponding
to Doppler velocities $\simeq$ 7000 km s${-1}$, and for the 1.78 $\AA$ (6.97 keV) line 
we can exclude Doppler velocities less than 430 km s$^{-1}$ and greater than 3900 km s$^{-1}$.  

\subsubsection{RGS}

Our procedure for fitting to the XMM RGS spectra is similar to that
used for the Chandra HETG.  One difference is that xspec encounters numerical problems when using the 
C statistic with RGS data, so we have rebinned the RGS data using the ftools tool grppha in order to 
have at least 20 counts per bin, and then used $\chi^2$ statistics. 
The results of fitting to various models are
shown in Table 6.  Figure 7 shows the ratio of the data to the model for the simplest
fit, that of a single power law plus neutral 
absorption.  The spectra have been binned for the purpose of plotting only.
The power law index has been forced to the value found by the fits to the MOS 
detectors.   For this fit $\chi^2/\nu$ is 1522/1672.  Figure 7 shows that there are significant residuals 
near 22, 19, 16 12, 10.6, 10., and 8.6 $\AA$ (0.59, 0.65, 0.78,
1, 1.2, and 1.44 keV),  which correspond to emission lines due to O VII 1-2, O VIII L$\alpha$, 
O VIII L$\beta$, Ne X L$\alpha$,  Ne X L$\beta$, and Mg XII L$\alpha$, respectively.  
In addition there are significant features at 22.9 and 23.1 $\AA$ 
which do not have secure identifications.
The parameters of a fit to the spectrum with Gaussian emission lines added at these 
energies is shown  in the second column of Table 6.  Inclusion of the narrow Gaussian 
lines results in a reduction 
in $\chi^2/\nu$ to 1392/1684.  Also given in table 6 are the line IDs, and the log of the 
probability of random occurrence according to an F test.
The spectral resolution of the RGS results in upper limits on the 
widths of the Gaussians which are greater than for the Chandra HETG. 
A notable difference between the Chandra spectrum and that shown here is the presence of the 
He-like O VII lines at 21.8 $\AA$ and 22.1 $\AA$ (0.569 and 0.561 keV).  These correspond to the 
intercombination and forbidden lines, respectively, and their best-fit intensity ratio is 1.3.
From this we infer that the predominant emission mechanism is recombination, rather than 
collisional excitation or resonance scattering, since either of these mechanisms is 
expected to produce strong permitted emission at 21.6 $\AA$ (0.574 keV).  In addition, the 
comparable strengths of the forbidden and intercombination components implies a density in the 
range 10$^9$ -- 10$^{11}$ cm$^{-3}$ (eg. Bautista and Kallman 1999; Porquet and Dubau 1999).
At greater densities the forbidden component is expected to be collisionally deexcited, while
at lower densities the intercombination line is weaker owing to the more efficient cascades to the
upper level of the forbidden line. 

Table 7 and  Figure 8 show the result of fitting the RGS data to an xstar  model.  
We again force the continuum slope to match that of the MOS spectrum.  
We fit to a two component model, one component with log($\xi$)=1 and one with log($\xi$)=4.5,
in order to simultaneously fit the low and high ionization lines in the RGS spectrum.
The statistical quality of the fit is 
significantly better than for the Gaussian fits, $\chi^2/\nu$=1418/1670.
This fit differs from the results of the xstar fits to the Chandra 
HETG spectra, for which we found an acceptable fit to a single ionization parameter, log($\xi$)=4.5.
An indicator of the difference is the O VII 1-2 line near 21 $\AA$
(0.59 keV), since this line requires lower ionization parameter in order to be 
emitted efficiently.  Figure 8 shows that this 2 component model adequately accounts for the 
spectrum in the 15-30 $\AA$ (0.4 -- 0.8 keV) range, but that there are inconsistencies between the model 
and the data in 8 - 15 $\AA$ (0.8 -- 1.6 keV) range.  Most notable is the feature near 9.7 $\AA$ 
(1.28 keV) which may be identified with Ne X L$\beta$.  This feature has a strength which is comparable
to the Ne X L$\alpha$ line at 10.2 $\AA$ (1.22 keV) and which is greater than can be produced by 
recombination or collisional excitation.    The narrow features near 10.8, 11.25, 11.45, and 11.7 $\AA$
(1.148, 1.110, 1.083, and 1.060 keV) are not statistically significant, and so are 
not fitted by the xstar model.  It is also worth noting the Ne IX 1-2 lines 
near 13.5 - 13.7 $\AA$ (0.9 -- 0.92 keV) which appear to be consistent with the predictions of the 
xstar model.  The best-fit abundances for 
this model are: O:Ne:Mg=0.3:1.6:2.1

The differences in the best-fit ionization parameters found from fitting to Chandra and XMM 
can be entirely attributed to the differing bandpasses of the two instruments.  The Chandra HETG 
detects such high ionization lines as S XVI and Si XIV L$\alpha$, which are outside the bandpass 
of the XMM RGS and which are only emitted efficiently at high ionization parameter.  
The RGS detects the O VII 1-2 lines which are outside the Chandra bandpass and which require 
a lower ionization parameter for efficient emission.  We have verified that this is the 
origin of the discrepancy by fitting both instruments to only the region of wavelength for which 
both have significant sensitivity, 7 -- 14 $\AA$ (0.88 -- 1.77 keV).  The result is that both fit 
equally well over a broad range of ionization parameter, since both the high ionization and low 
ionization lines have been excluded in this fit.

\section{Discussion}

\subsection{Velocity Offsets and Widths}

In figure 9  we compare the velocity offsets of the lines observed by the various 
instruments.  The velocity
offsets are determined using line wavelengths from the NIST database, 
where available, and from the compilation of Verner (1999) for the others.
Figure 9 shows that essentially all the line wavelengths are consistent 
with being emitted at zero velocity, although the uncertainties 
are typically $\sim$1000 km/s or greater.  Only the wavelength of the Fe XXVI
L$\alpha$ near 1.78 $\AA$ (6.97 keV) as measured by the Chandra HETG is inconsistent with  
zero velocity, although the wavelength of this line as measured by the 
XMM PN is consistent with zero velocity. In the case of the line observed at 
1.85 $\AA$ (6.7 keV) we include the data for all 3 of 
the components of the 1-2 lines of Fe XXV, with the result  that 
the wavelengths of the forbidden and semi-forbidden lines are consistent with 
smaller velocity offsets than the allows line.  This suggests that, if 
the line-emitting material does not have a high bulk velocity relative to 
us, then the lines are formed predominantly by recombination.

The limits on line widths allow us to approximately constrain the 
location of the line emitting gas, under the assumption that the 
gas motions are Keplerian in the vicinity of a 1 $M_\odot$ compact object.
If so, the limits on the velocity obtained for the 
1.78 $\AA$ (6.97 keV) line using the XMM PN correspond to a radii between 8.8 $\times 10^8$ cm
and 7.2 $\times 10^{10}$ cm.  Although there are no 
independent measures of the mass, it is likely that the inclination $i$ is 
close to 90$^o$.  This can be compared with estimates for the 
disk outer radius.  Using standard expressions for the Roche lobe radii (eg. Frank, King and 
Raine, 1992), and 
assuming values for the mass ratio $q$ in the range 1 -- 2.2 (Shabhaz et al., 1999) the 
radius of the compact object Roche lobe is in the range 0.9 -- 1.2 $\times 10^{12}$ cm.
Although estimates for the disk radius itself are less certain, it is clear 
that the iron emission originates at a small fraction 
($\sim$ 10$\%$) of the probable outer disk radius.  
The width of the 1.94 $\AA$ (6.4 keV) iron K$\alpha$ line measured by the 
XMM PN is much greater, 0.27 keV.  
This  corresponds  to the Keplerian broadening at a disk radius of $7 \times 10^7$ cm. 
This difference in widths is surprising if it actually reflects the range of 
radii where the lines are emitted.  A more likely explanation is that the width 
of the 1.94 $\AA$ (6.4 keV) line is due to blending of multiple components of the 
fluorescence  lines from various ionization stages of iron.  If so, the estimate 
of radius obtained from the 1.78 $\AA$ (6.97 keV) line remains relevant.

\subsection{Line Luminosities, Variability, and Emission Measures}

The fluxes we derive for the various experiments are 5.4, 3.8, and 5.9  $\times 10^{-11}$
erg cm$^{-2}$ s$^{-1}$ in the 2-10 keV band for ASCA, Chandra, and XMM respectively.
The fact that the ASCA and XMM are consistent, and that the Chandra flux is slightly 
lower, further confirms the conclusion that the fact that the Chandra spectrum requires 
only a high ionization parameter component while the XMM RGS requires both high ad low ionization 
parameter components is not due to variability.  The lower ionization parameter inferred from the XMM fit
would suggest a lower flux during the XMM observation if variability were the 
origin for the difference.  However, the difference is in the opposite direction, and the Chandra
observation has lower flux.  From this we conclude that the difference in the best fit 
ionization parameter for the two observations is due to the differing wavelength bands of the 
two instruments together with an intrinsic distribution in ionization parameter within the 
source, rather than variability.

Figure 11 shows the emissivities for the various lines we observe as a
function of ionization parameter, $\xi$, as calculated for a library of 
single-zone xstar photoionization models.  Also included in this are lines which we 
do not detect, and whose absence places constraints on the conditions in the 
ADC.  These include principally the lines of He-like ions such as 
Ne IX, Mg XII, and Si XIII.  From this figure it is clear that 
the absence of these lines in the Chandra HETG spectra implies 
ionization parameter values greater than $\xi \sim 10^4$.  Conversely, 
the best fit to the XMM RGS spectrum that we find is affected by the 
need to fit the O VII line near 21 $\AA$, and this requires ionization parameters
$\xi \sim 10^1$.  

The line strengths can be used to infer the emission measure of the gas 
in 2S0921-630 by dividing the line luminosity by the emissivity of the various
lines.  As example of such a calculation is shown in figure 10, for the 
various strong lines detected by Chandra and XMM.  These estimates
were derived by using the line emissivities shown in figure 11, and summing over 
an arbitrary distribution of ionization parameters binned into 9 bins equally spaced 
in log($\xi$) from 1 to 5.  The result is a model luminosity for each line, 
$L^{mod}_i=\Sigma_{\xi_k} j_{i,k} EM_k$, where $ j_{i,k}$ is the emissivity for line $i$ in 
ionization parameter bin $k$ as shown in figure 11 and $EM_k$ is the emission measure distribution.
This was fit to the observed luminosities and errors $L^{obs}_i$ and $\sigma^{obs}_i$, 
derived from tables 2, 5 and 6 using a distance of 7 kpc by 
varying the emission measure in each bin independently until a minimum 
was found in the figure of merit.  We 
take the figure of merit to be to be $\Sigma_i ((L^{mod}_i-L^{obs}_i)/\sigma^{obs}_i)^2$.  This 
procedure is not expected to be highly accurate, owing to the 
simplified physical assumptions and the 
coarseness of the various grids, these results in the upper panel of figure 10 show that 
we are able to fit most line strengths to within a factor of 2.
The lower panel of figure 10
 shows the derived best-fit emission measure distribution, which corresponds to 
3 distinct components.  This is similar to the result obtained by fitting xstar models 
to the Chandra and XMM data separately.  This distribution is not consistent with 
a power law or other smoothly varying function.  

 An equivalent procedure is to use the normalization 
calculated using the fit to the xstar model.   The general procedure for 
doing this is outlined in the xstar manual and in the Appendix to this paper.
The result is $\sim 10^{57}$ cm$^{-3}$, which can be compared with that expected 
from an ADC.

Estimates for the emission measure of an ADC depend on several  quantities
which are not accurately determined, including the geometry of the corona 
and its effect on the transfer of radiation, the intrinsic luminosity and 
spectrum of the continuum X-ray source, the possible role of outflows or
winds, and the presence of any other heating or ionization mechanism such 
as magnetohydrodynamic processes (eg. Stone and Miller, 2001).
Simple estimates can be obtained by assuming the corona is hydrostatic and 
unaffected by MHD processes, as was done by Kallman and White (1989).
If so, the corona can be crudely represented as a series of concentric 
cylinders, each having a Gaussian density distribution with height, and where 
the local scale height is:

$$z_s(R)=\sqrt{{kT R^3}\over{G M m_H}}$$

\noindent where T is the local gas temperature and M is the mass 
of the central compact object.  The maximum density in the corona is 
attained when the ionization parameter is $\Xi=\Xi_H^*\simeq$1--10, 
(eg. Krolik McKee and Tarter, 1981), where $\Xi=L/(4 \pi R^2 nckT)$.
Then the emission measure is approximately

$$EM(R)\simeq n_{max}^2 4 \pi R^2 z_s 
= 1.2 \times 10^{61} L_{38}^2 R_{10}^{-1/2} T_7^{-3/2} \Xi^{-1}
 {\rm cm}^{-3}$$

\noindent where $T_7$ is the temperature in units of 10$^7$K, $R_{10}$
is the radius in units of 10$^{10}$ cm, and $L_{38}$ is the continuum 
luminosity in units of 10$^{38}$ erg s$^{-1}$.  This shows that the 
emission measures inferred from the photoionization model fits can be 
attained by an accretion disk corona for a plausible range of parameters, 
i.e. $L_{38} \sim 10^{-2}$, and all other parameters of order unity.

A further constraint on the ADC comes from the derived density in the 
O VII lines.  If the intrinsic luminosity of the X-ray source is 
10$^{38}$ erg s$^{-1}$, then the minimum distance at which material at the 
inferred O VII maximum density of 10$^{11}$ cm$^{-3}$ and ionization parameter
log($\xi)\simeq$2 can exist under photoionization is 2.2 $\times$ 10$^{12}$ cm.
This is larger than the likely outer radius of the accretion disk, which is 
less than $\sim$10$^{12}$ cm (see below).  This estimate assumes 
that the local radiation intensity 
in the ADC is determined by geometrical dilution of the 
X-rays from the compact 
object.  The most likely explanation is that the intensity of the ionizing continuum radiation
within the O VII region of the ADC is reduced by scattering or absorption.  
If so, the amount of attenuation required in order to reduce the O VII radius to 
a value consistent with the disk is a factor $\sim$100.  This differs 
from the simple ADC models which have been constructed by eg. Ko and Kallman (1996)
and Jimenez-Garate et al. (2001), 
in which ionization parameters low enough to produce O VII could only occur at 
densities $\sim 10^{14}$ cm$^{-3}$ or greater.  An alternative explanation is that 
the X-ray source is intrinsically faint, $\sim 10^{36}$ erg s$^{-1}$ or less, or 
that the O VII emission region lies outside the traditional estimates for the disk 
outer radius.

\subsection{Comparison with other LMXBs}

Although 2S0921-630 is not the brightest ADC source, it may the 
the most suitable target for studying the accretion disk corona 
during eclipse.  This is based on the ASCA spectrum, and is 
likely due to its long orbital period, which allows
long observations to be carried out entirely during eclipse.
An illustration of this is provided by the spectrum of the 
ADC source 4U1822-371 (Cottam et al., 2001), which shows many of the same features as 
0921-63.  However, 
the line emission from 1822-371 is much weaker in spite of its 
greater total flux.  In addition, in 4U1822 the emission lines don't come 
from the corona.  They are phase-dependent and appear to come from the 
photo-illuminated bulge at the impact point.  This suggests that in 4U1822-371
there may be more dilution of the line emission by directly observed 
X-rays than is expected, or perhaps the inclination is such 
that we detect more direct X-rays than in 2S0921-630.

\subsection{Neutral-Like Iron line}

The iron K line is unique among X-ray lines from abundant elements 
in that it can be emitted efficiently by neutral or nearly-neutral gas
via inner shell fluorescence in addition to being emitted efficiently 
when iron is highly ionized (here and in what follows we define 'neutral-like' 
to be ion stages less than Fe XVII, so that the line energy is indistinguishable
from neutral).  Although many LMXBs (including 2S0921-630) show 
emission from highly ionized iron, a uniformly illuminated accretion disk is expected to 
radiate a strong line from nearly neutral iron.  The strength of this line 
is a constraint on the strength of illumination of the accretion disk 
at radii $\geq 10^8$cm, where the iron in the disk photosphere is expected to 
neutral-like.  A simple estimate for the flux  of a neutral-like fluorescence line 
emitted locally from an illuminated disk is:

$$F_{fluorescence}(R)=\int_{\varepsilon_{Th}}^{\infty}{F^{continuum}_{\varepsilon}(R)d\varepsilon} 
(1-e^{-\tau_{Fe K}}) \omega$$

\noindent where $\varepsilon_{Th}$ is the threshold energy for photoionization, 
 $\tau_{Fe K}$ is the optical depth at this energy and is assumed to be large, 
$F^{continuum}_{\varepsilon}(R)$
is the continuum flux,  $\omega$ is the fluorescence yield and $y_{Fe}$ is the iron abundance 
relative to solar (Grevesse et al., 1996).  If we define 
the incident continuum flux in terms of the luminosity from the neutron star, $L$, 
the radius $R$, a factor $f\leq$1 which accounts for non-normal incidence, attenuation between the 
continuum source and the disk, etc., and a factor $\kappa$ which is the fraction 
of the continuum flux absorbed by iron K, we can write:

$$\kappa f(R) {{L}\over{4 \pi R^2}}=\int_{\varepsilon_{Th}}^{\infty}{F^{continuum}_{\varepsilon}(R)d\varepsilon} $$

\noindent Then the total fluorescent line luminosity is obtained by integrating over the 
disk surface between two radii $R_{min}$ and $R_{max}$:

$$L_{fluorescence}=\omega y_{Fe} \kappa {{L}\over{2}} \int_{R_{min}}^{R_{max}} {{f(R)}\over{R}} dR$$

\noindent If we take $f(R)$=constant, then 

$$L_{fluorescence}\simeq \omega y_{Fe} \kappa {{L}\over{2}} {\rm ln}({R_{max}}/{R_{min}}) f$$

\noindent Taking plausible values for these quantities: $\omega$=0.34 (valid for neutral-like iron), 
$y_{Fe}=1$, $\kappa$=0.2 (for an $\varepsilon^{-1}$ power law)
$L=3 \times 10^{35}$ erg s$^{-1}$ (the measured value for a distance of 7 kpc),  
${R_{max}}/{R_{min}}$=100 (an upper limit with little effect on the result),
$f$=0.1 (an upper limit; cf. Ko and Kallman 1994), we get $L_{fluorescence}\simeq 5 \times 10^{33}$ erg s$^{-1}$.
This is greater than the upper limit allowed by the observations for the 6.4 keV line, 
$1.8 \times 10^{33}$ erg s$^{-1}$, suggesting that the illumination is weaker than we have
assumed, $f\leq 0.03$, or that the iron abundance is less than solar.

\section{Summary}

The complementary capabilities of Chandra and XMM provide new insight into the 
reprocessing gas in 2S0921-630.  The XMM RGS allows detection of 
line emission from elements such as oxygen, while the Chandra HETG allows detection of
the K lines of sulfur and iron.  The XMM EPIC detectors allow measurement of  the 
spectrum above 7 keV and measurements of line widths.  
We have found line emission from medium-Z elements, O, Ne, Me, Si, S, and 
also K lines from iron.  With the exception of the neutral-like and H-like components 
of the Fe K line as measured by the PN detector on XMM, the lines have widths which are 
less than can be measured with the Chandra and XMM detectors.  The measured
widths for the two components of the iron line suggest upper limits to the size of the 
ADC of 7$\times 10^{10}$ cm for a 1 M$\odot$ compact object at 90$^o$ inclination 
if orbital Doppler broadening is the dominant mechanism.  The 
line spectra from each grating instrument fit adequately with photoionized models.
However, the XMM observations require a second component at lower ionization parameter
than  inferred for Chandra.  
The O VII lines are detected in the XMM RGS and imply gas densities in the range 
10$^9$ -- 10$^{11}$ cm$^{-3}$, which in turn implies that either the X-ray source is 
intrinsically faint, that the emission region lies outside of the 
accretion disk, or that large regions of the ADC are very optically thick.
The distribution of emission measures of xstar models which comes closest to fitting 
the ensemble of the line 
data is a 3 component model, similar to that inferred from the separate xstar fits 
to the Chandra and XMM grating data.
The emission measures inferred from both a simple optically thin DEM analysis and from 
the xstar fits are consistent with simple estimates for 
an ADC if the intrinsic luminosity of the source is approximately what we observe.  
The high energy spectrum as observed by the PN shows complex structure at energies above 5 keV:
curvature in the continuum spectrum, 
multiple iron K$\alpha$ components similar to those seen in the Chandra HETG, 
plus either additional emission features at 7.8 keV and 8.2 keV, absorption,
or reprocessed emission which differs from standard reflection models.

The picture of the ADC that we are left with is as follows:  The total ADC emission measure 
is comparable to that obtained from simple estimates, although the distribution of ionization 
parameter is not smooth, monotonic, or easily predicted or compared with models.  
The width of the Fe XXVI L$\alpha$ line implies that high ionization gas is 
distributed primarily at radii less than $\simeq 7 \times 10^{10}$cm.  The intensity
of the neutral-like Fe K$\alpha$ line implies that this material is primarily at large radii, 
i.e. greater than $\simeq 10^9$cm.  If so, the measured width of this line is due to blending 
of multiple components due to different ion stages.  The constraints on the density of the O VII 
emitting gas require that either the intrinsic luminosity of the continuum source be low, i.e. 
$\leq 10^{36}$ erg s$^{-1}$, or that the O VII emitting region be heavily shielded 
from direct X-rays.  The first scenario implies that we observe the full luminosity 
of the continuum source.  This conflicts with ADC models in which the 
intrinsic luminosity of ADC source is much greater than we observe and that much of what we 
observe is scattered or diffused through the ADC.  The second scenario requires a 
large column density ($\geq 10^{23}$ cm$^{-2}$) of absorbing or scattering gas which is not 
apparent from the observed spectrum.  The high energy spectrum, above the iron K line region, 
shows evidence for complicated and unexplained emission, or possibly absorption.  Although simple 
models for absorption do not fit the spectrum as well, both emission and absorption models 
produce fits which are statistically acceptable.  The absorption scenario is more 
plausible owing to the 
likelihood of dense gas associated with the disk atmosphere, bulge, or wind impinging on the line 
of sight.  If so, this gas is highly ionized and also has large column density 
($\geq 10^{24}$ cm$^{-2}$).  More sensitive observations, or more 
secure EPIC calibration, in the 1-20 keV energy band may allow this 
to be tested; an absorption scenario would predict that the underlying power law would 
recover at energies $\sim$15 keV.

\noindent \underbar{Acknowledgements} This work was supported by grants 
from NASA through the Chandra and XMM guest observer programs.

\references

Arnaud, K.A., 1996, Astronomical Data Analysis Software and Systems
                     V, eds. Jacoby G. and Barnes J., p17, ASP Conf. Series volume 101\par

Asai, K., Dotani, T., Nagase, F., and Mitsuda, K., 2000, Ap. J. Supp., 131, 571\par

Bautista, M., and Kallman, T., 2000, ApJ, 544, 581\par

Bautista, M., and Kallman, T., 2001, Ap. J. Supp., 133,221\par

Begelman, M., McKee, C., and Shields, 1982, Ap. J., 271, 70\par

Branduardi-Raymont, G., et al., 1981, Space Sci. Rev., 30, 279\par

Branduardi-Raymont, G., et al., 1981, MNRAS, 205, 403\par

Chevalier, C., and Ilovaisky, S., 1981, Astr. and Ap., 94, L3\par

Chevalier, C., and Ilovaisky, S., 1982, Astr. and Ap., 112, 68\par

Cottam, J., et al.,  2001 ApJ, 557, 101\par
 
Cowley, A., et al., 1982, Ap. J., 256, 605\par

Frank, J., King, A., and Raine, D., 1992 'Accretion Power in Astrophysics', Cambridge Astrophysics 
Series (Cambidge University Press; Cambridge)\par

Gottwald, M ,et al. 1995, A. and A. Supp., 109, 9\par

Grevesse, N., Noels, A., and Sauval, A. 1996, in ASP Conf. Ser. 99, Cosmic Abundances, 
ed. S. Holt and G. Sonneborn (San Francisco: ASP), 117\par

Jimenez-Garate, M., et al., 2001, ApJ, 558, 448\par

Kallman, T., and Bautista, M., 2001, ApJS, 133, 221\par

Kallman, T., and White, N., 1989, Ap. J., 341, 955\par

Kallman, T., Liedahl, D., Osterheld, A., Goldstein, W., and Kahn, S., 
1996, Ap., J., 465, 994\par

Ko, Y., and Kallman, T., 1994, Ap. J., 431, 273\par

Krolik, J., McKee, C., and Tarter, C. B., 1981, Ap. J. 249, 422

Mason, K., et al., 1987, MNRAS, 226, 423\par

Mason, K. O., 1986, in, Physics of Accretion onto Compact Objects,, p.29,
eds Mason, Watson, and White, Springer-Verlag, Heidelberg \par

McClintock, J., et al., 1982, Ap. J., 258, 245\par

Porquet, D., and Dubau, J.,  2000, A and AS, 143, 495\par

Proga, D., and Kallman, T.,  2002, ApJ, 565, 455\par

Shahbaz, T., et al., 1999, Ast. Ap., 344,101\par

Vrtilek, S., Soker, N., and Raymond, J., 1994, Ap. J., 404, 696\par 

Verner, D. A., Verner,  E. M. and  Ferland, G. J., 1996, 
Atomic Data Nucl. Data Tables, 64, 1 \par

White, N.E., and Holt, S., 1982, Ap. J., 257, 318 \par

White, N.E., 1996, in, X-Ray Imaging and Spectroscopy of Cosmic 
Hot Plasmas,, ed. F. Makino\par

Woods, D., et al.,  1996, ApJ, 461, 767\par

\newpage

\centerline{Appendix}

The general procedure for inferring the emission measure from the normalization 
calculated using the fit to the xstar model is as follows. 
The observational data is fitted to a library of xstar table models 
constructed using the xstar2xspec tool.  When this library was calculated 
a source luminosity $L_{xstar}$ was used, and when the data was fitted to the 
model a best fit normalization, xstar ionization parameter ($\xi$) and 
column density ($N$) were derived.  The best-fit xstar model was calculated 
as a spherical shell, and its  emission measure is 

$$EM_{xstar}= n^2 4 \pi R^2 \Delta R = 4 \pi {{L_{xstar} N}\over{\xi}}$$

\noindent If the distance to the source ($D$) and its intrinsic luminosity ($L$) are known, 
then the emission measure of the source is given by

$$EM={\rm normalization} \times EM_{xstar} \left({{L_{xstar}}\over{L}} 4 \pi D_{kpc}^2 \right)$$

\noindent where $D_{kpc}$ is the distance to the source in kpc.

   In the case of the 
fit to the Chandra HETG spectrum the best fit normalization is 210, for 
an ionization parameter $\xi \simeq$ 10$^5$, and $L_{xstar}=10^6$ and $EM_{xstar}=2 \times 10^{61}$.  
This corresponds to a source  emission measure
of $EM=6.8 \times 10^{54}$ cm$^{-3}$ L$_{38}^{-1}$ D$_7^{2}$, where L$_{38}$ is 
the continuum luminosity in units of 10$^{38}$ erg s$^{-1}$ and D$_7$ is 
the distance in units of 7 kpc.  

\newpage

\begin{figure}[tb]
\plotfiddle{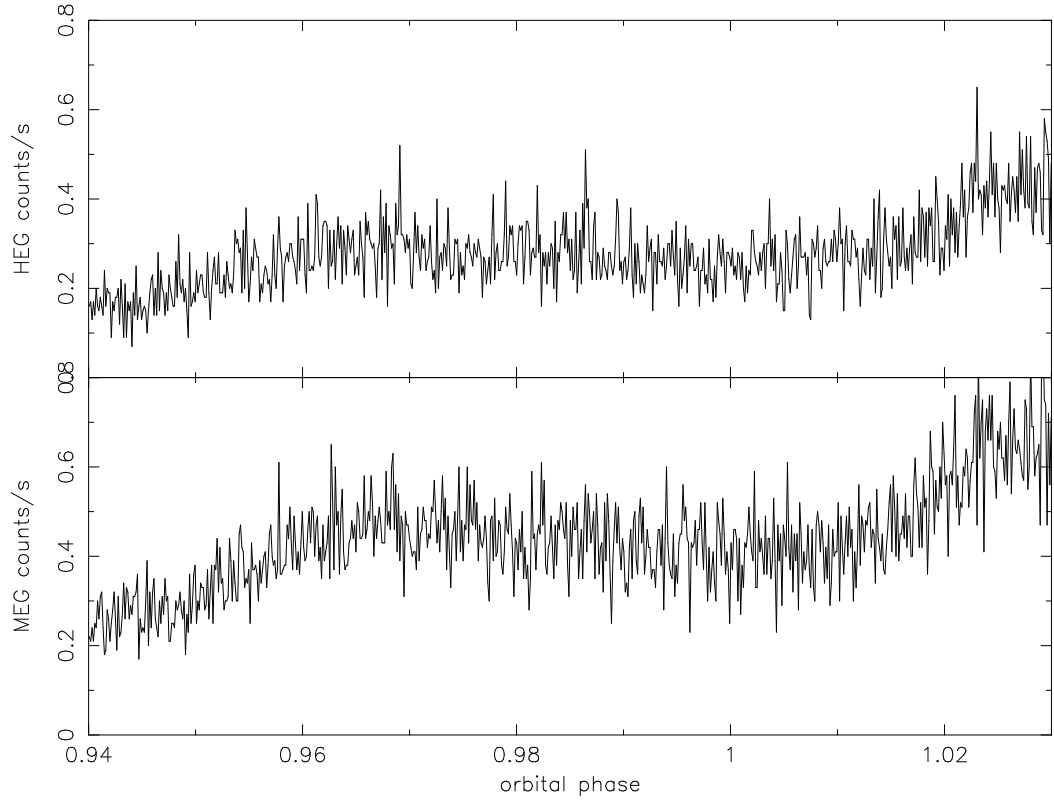}{12cm}{-90}{60}{60}{-250}{400}
\caption{HEG light curve during the Chandra Observation of 2S0921-63
plotted vs. orbital phase in 100 second time bins}
\end{figure}

\begin{figure}[tb]
\plotfiddle{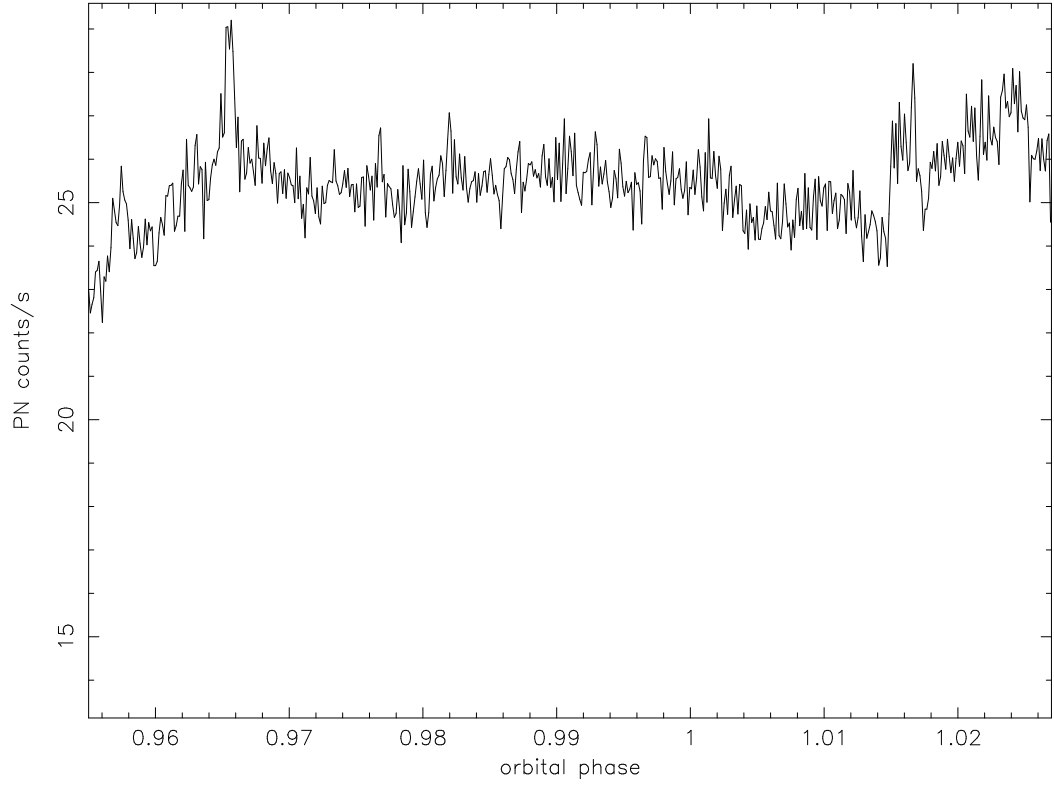}{12cm}{-90}{60}{60}{-250}{400}
\caption{PN Light curve during the XMM Observation of 2S0921-63 plotted vs. 
orbital phase in 1 second time bins.}
\end{figure}

\begin{figure}[!bp]
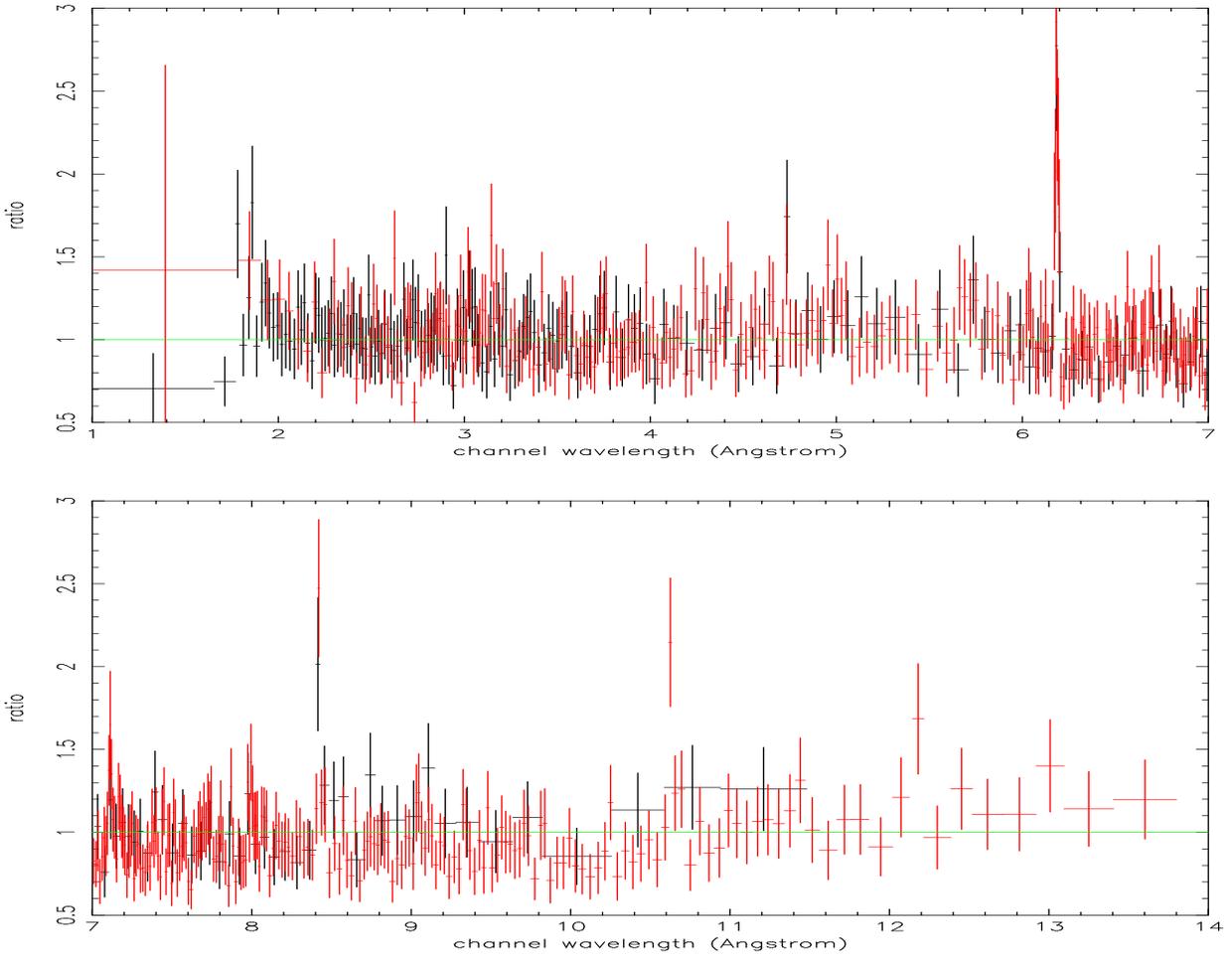

\plotfiddle{f3a.ps}{6cm}{270}{70}{35}{-260}{200}
\plotfiddle{f3b.ps}{6cm}{270}{70}{35}{-260}{200}
\caption{Ratio of HETG spectrum to absorbed power law model.
HEG (read) and MEG (black) are displayed separately.}
\end{figure}

\begin{figure}[!bp]
\plotfiddle{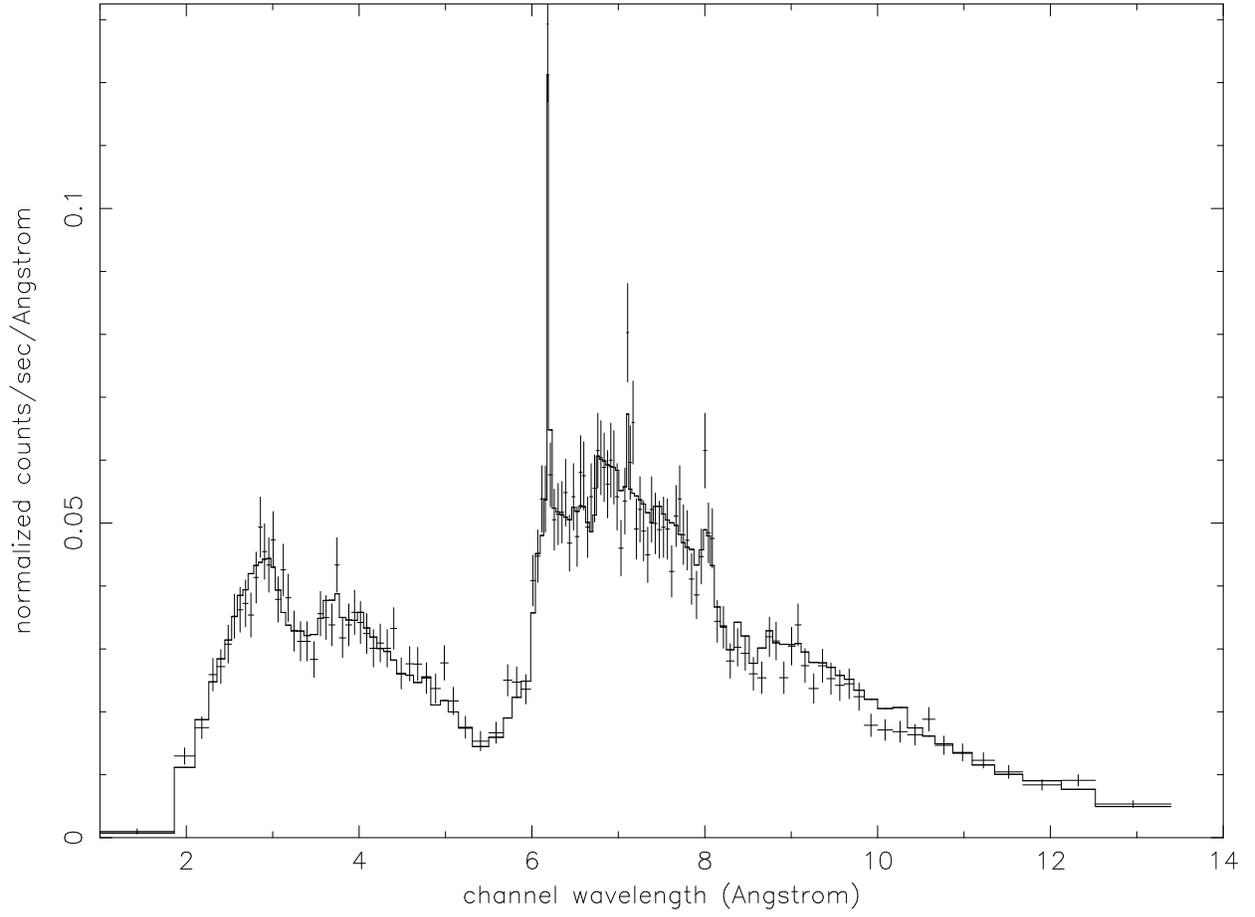}{12cm}{270}{70}{70}{-260}{400}
\caption{Xstar model fit to HETG spectrum.  Only the HEG is shown for clarity.}
\end{figure}

\begin{figure}[tb]
\plotfiddle{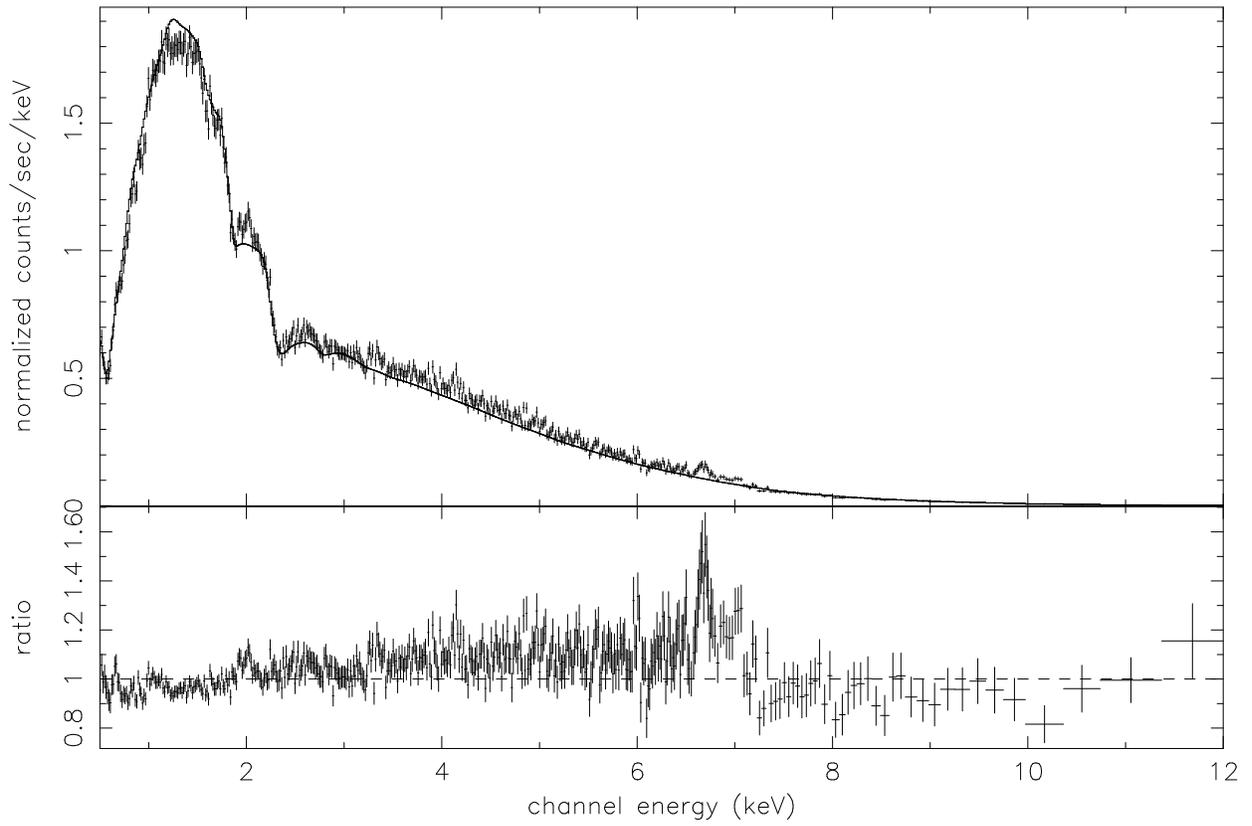}{12cm}{270}{70}{70}{-260}{400}
\caption{EPIC-MOS spectrum, counts + model (upper panel) and 
counts/model ratio (lower panel).  Only MOS 1 is shown for clarity.}
\end{figure}

\begin{figure}[tb]
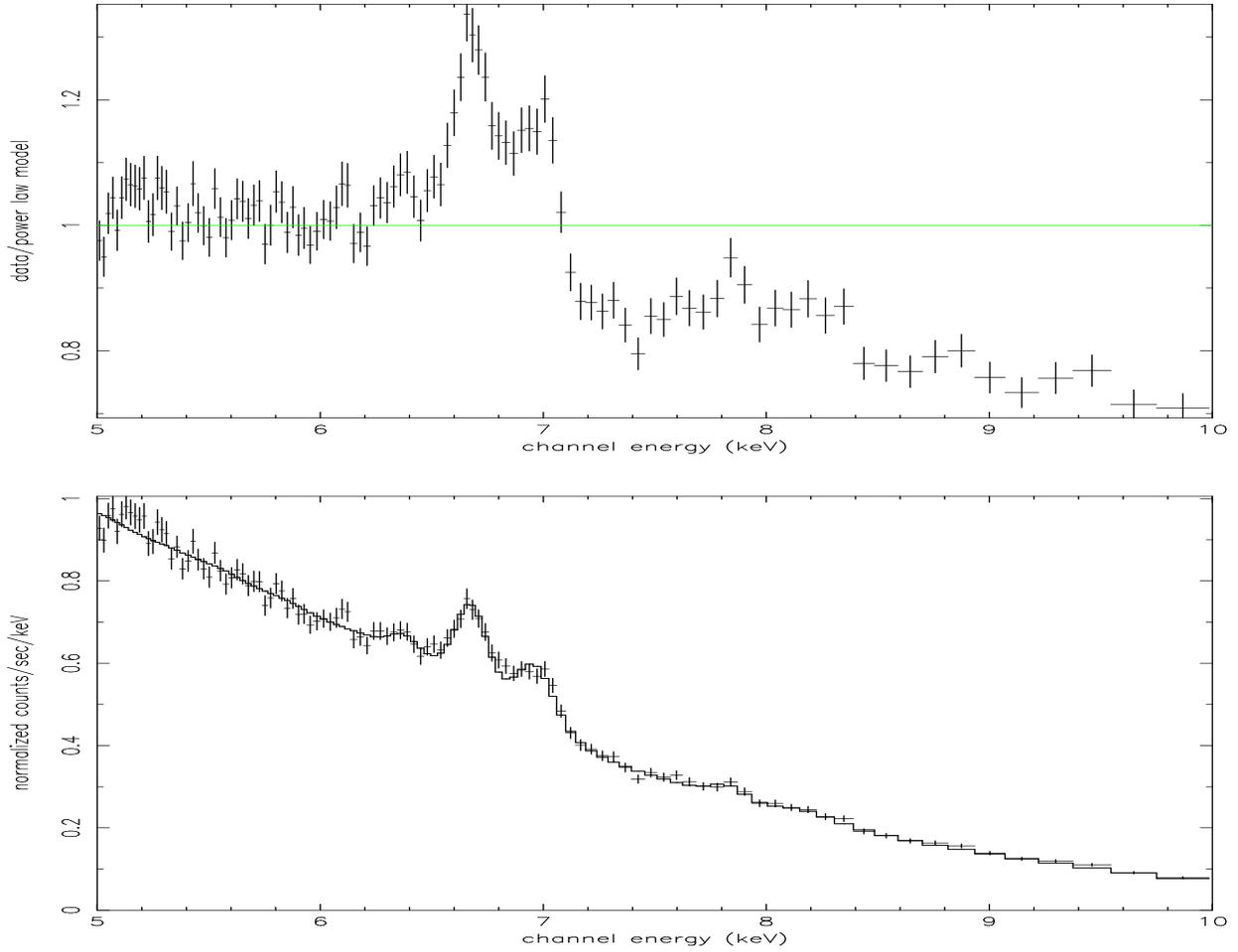

\plotfiddle{f6a.ps}{6cm}{270}{70}{35}{-260}{200}
\plotfiddle{f6b.ps}{6cm}{270}{70}{35}{-260}{200}
\caption{EPIC PN spectrum in the vicinity of the iron K line,
Upper panel: Ratio of data to cutoff power law model.
Lower panel: Fit to 
cutoff power law plus Gaussian emission lines and edge.}
\end{figure}

\begin{figure}[!bp]
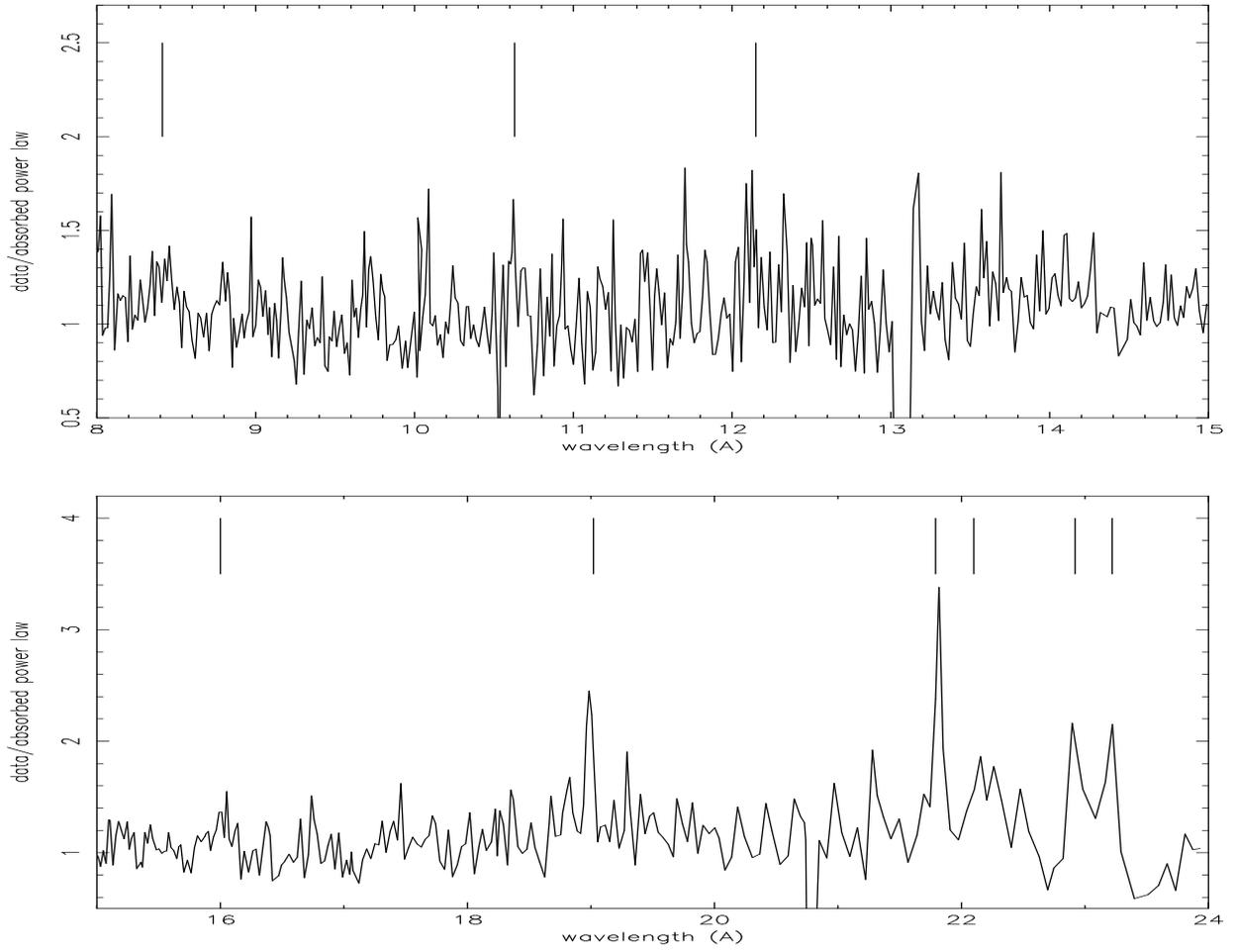

\plotfiddle{f7a.ps}{6cm}{270}{70}{35}{-260}{200}
\plotfiddle{f7b.ps}{6cm}{270}{70}{35}{-260}{200}
\caption{Ratio of RGS spectrum to absorbed power law model.  
Strong lines are marked with vertical bars. RGS 1 and RGS 2 have been 
combined}
\end{figure}

\begin{figure}[!bp]
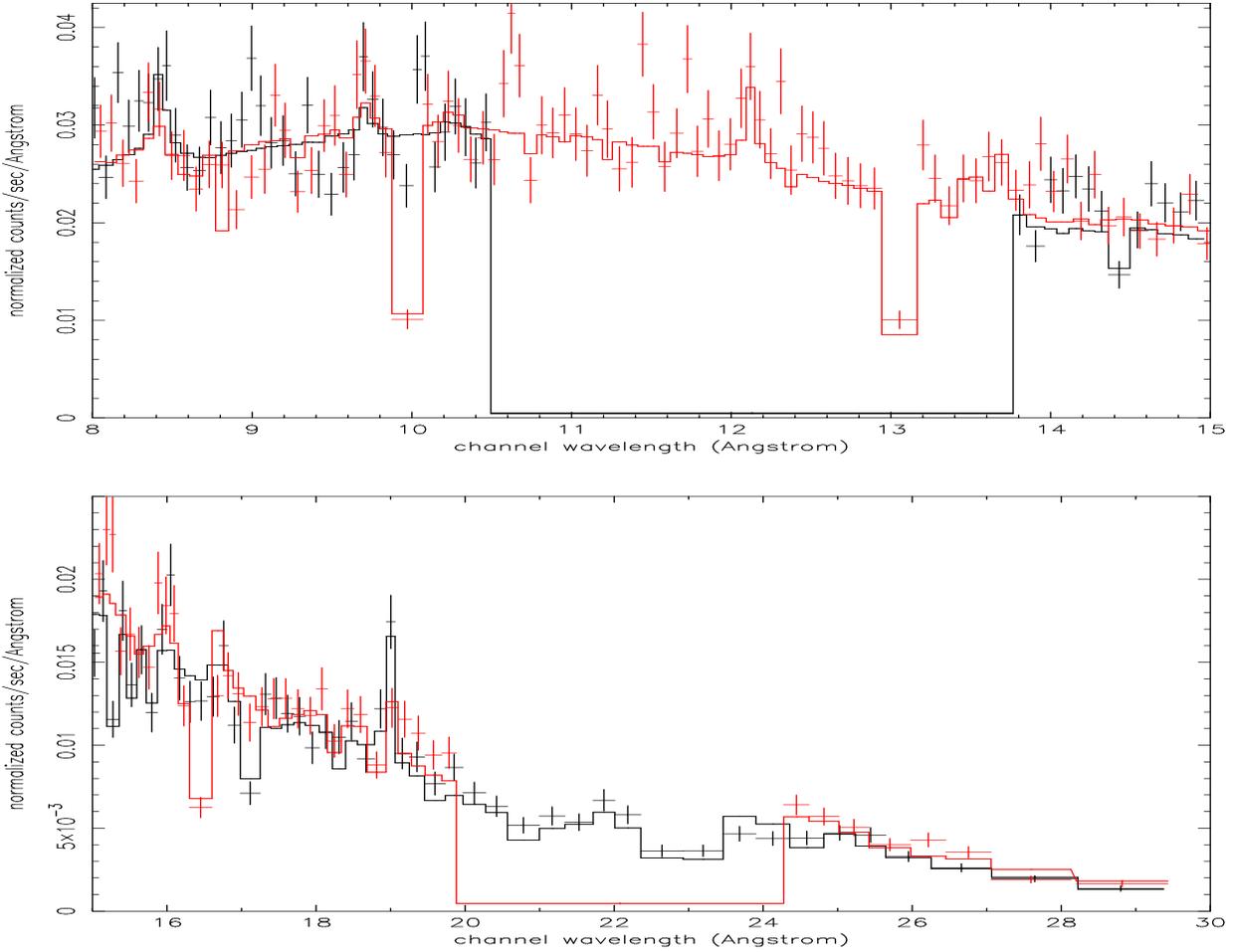

\plotfiddle{f8a.ps}{6cm}{270}{70}{35}{-260}{200}
\plotfiddle{f8b.ps}{6cm}{270}{70}{35}{-260}{200}
\caption{Xstar model fit to RGS spectrum using 2 component model 
as described in the text. RGS 1 and RGS 2 are plotted separately.
Upper panel:  8 -- 15 $\AA$ (0.8 -- 1.6 keV) range. 
Lower panel: 15 -- 30 $\AA$ (0.4 -- 0.8 keV) range.}
\end{figure}

\begin{figure}[tb]
\plotfiddle{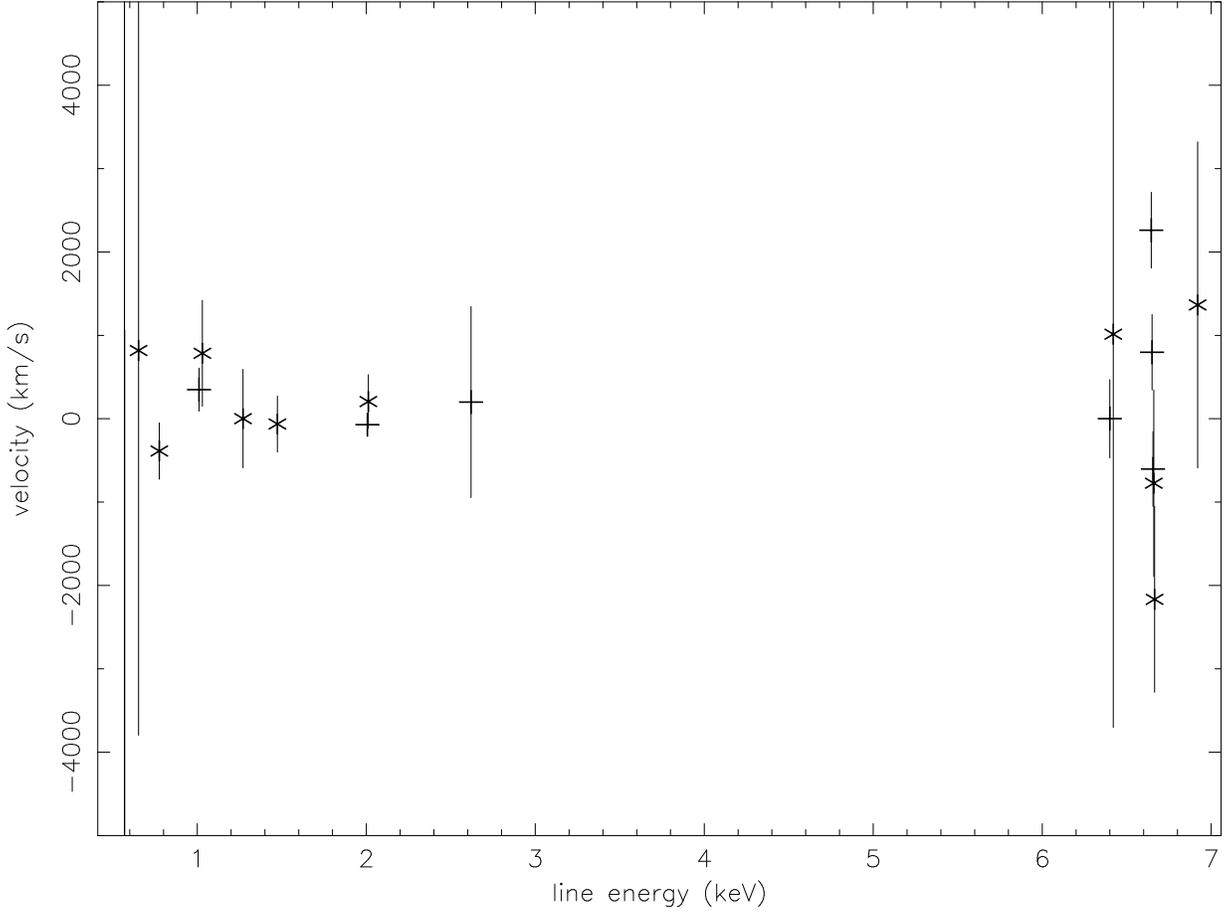}{10cm}{270}{70}{70}{-300}{400}
\caption{Comparison of line velocities derived from XMM (*) and Chandra (+)}
\end{figure}

\begin{figure}[tb]
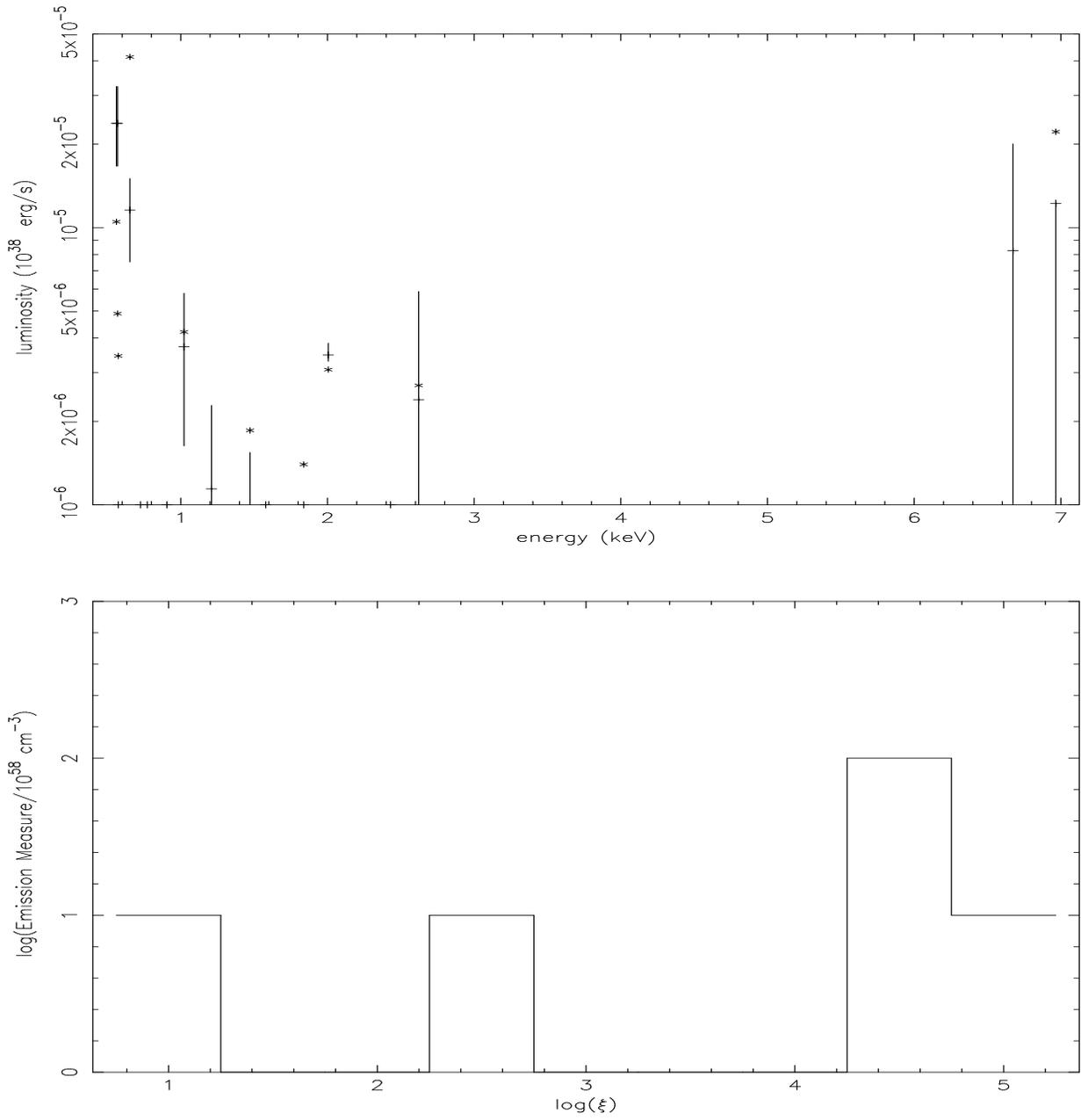

\plotfiddle{f10b.ps}{8cm}{270}{70}{45}{-300}{300}
\plotfiddle{f10a.ps}{8cm}{270}{70}{45}{-300}{300}
\caption{Upper panel: Comparison of line luminosities derived from XMM and Chandra (+)
with the result of a differential emission measure fit to the data using xstar models (*).  Lower panel:
best fit emission measure distribution}
\end{figure}

\begin{figure}[tb]
\plotfiddle{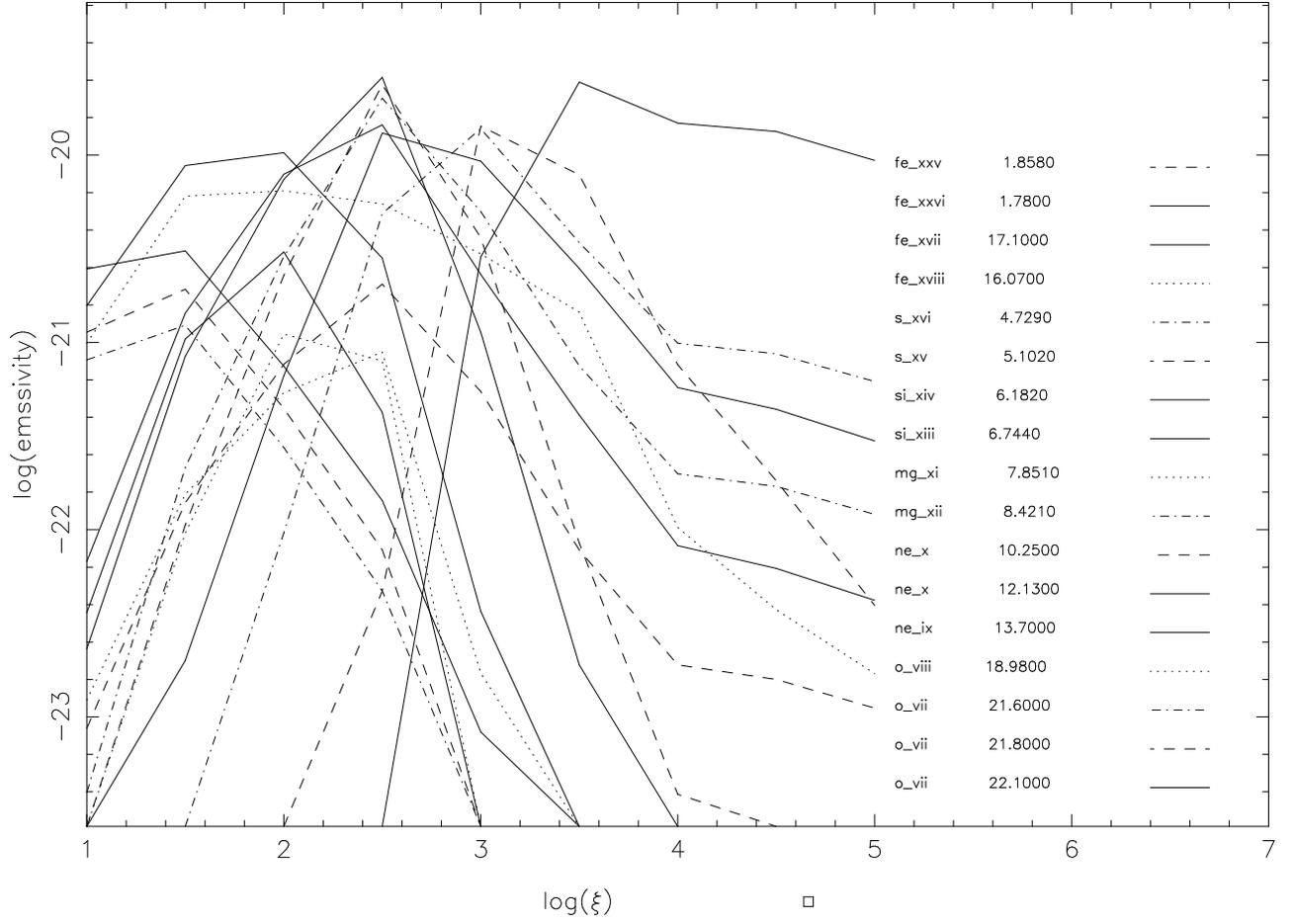}{10cm}{270}{70}{70}{-260}{400}
\caption{Emissivities in erg cm$^3$ s$^{-1}$ for strong lines in the Chandra/XMM band 
as calculated by xstar}
\end{figure}

\clearpage

\begin{deluxetable}{lrrrrcrrrrr}
\tablenum{1}
\tablecaption{ASCA Lines}
\tablehead{
\colhead{quantity} &  \colhead{value}}
\startdata
Absorption, NH ($\times 10^{22}$ cm$^{-2}$)&0.17\nl
Power Law Index (photons)&1.1\nl
Cutoff Energy (keV)&17.8\nl
normalization (erg cm$^{-2}$ s$^{-1}$) &4.0$^+_-$0.1 $\times 10^{-3}$\nl
Flux 2-11 keV (erg cm$^{-2}$ s$^{-1}$)&3.45 $\times 10^{-11}$\nl
line 1 &&\nl
Wavelength ($\AA$)&12.4\nl
Energy (keV)&1.0$^{+0.02}_{-0.75}$\nl
normalization ($\times 10^{-5}$erg cm$^{-2}$ s$^{-1}$) &9.7$^{+10.}_{-6.}$\\
Equivalent Width (eV)&22$^+_-$8\nl
Line ID &Ne X L$\alpha$\nl
log(P)&-4.3\nl
line 2 &&\nl
Wavelength ($\AA$)&10.7\nl
Energy (keV)&1.15$^{+0.05}_{-0.08}$\nl
normalization ($\times 10^{-5}$erg cm$^{-2}$ s$^{-1}$) &4.7$^{+5.}_{-4.}$\\
Equivalent Width (eV)&22$^+_-$8\nl
Line ID &Ne X L$\beta$\nl
log(P)&-3.0\nl
\enddata
\end{deluxetable}

\clearpage

\begin{deluxetable}{lrrrrcrrrrr}
\tablenum{1}
\tablewidth{400pt}
\tablecaption{ASCA Lines, continued}
\tablehead{
\colhead{quantity} &  \colhead{value}}
\startdata
line 3 &&\nl
Wavelength ($\AA$)&6.2\nl
Energy (keV)&1.99$^+_-$0.1\nl
normalization ($\times 10^{-5}$erg cm$^{-2}$ s$^{-1}$) &3.5$^{+3.}_{-2.}$\\
Equivalent Width (eV)&20$^+_-$11\nl
Line ID &Si XIV L$\alpha$\nl
log(P)&-2.2\nl
line 4 &&\nl
Wavelength ($\AA$)&4.77\nl
Energy (keV)&2.6$^+_-$0.1\nl
normalization ($\times 10^{-5}$erg cm$^{-2}$ s$^{-1}$) &1.5$^{+2}_{-2}$\\
Equivalent Width (eV)&15$^+_-$9\nl
Line ID &S XVI L$\alpha$\nl
log(P)&-0.32\nl
line 5 &&\nl
Wavelength ($\AA$)&1.82\nl
Energy (keV)&6.8$^+_-$0.1\nl
normalization ($\times 10^{-5}$erg cm$^{-2}$ s$^{-1}$) &4.7$^{+2.3}_{-3.7}$\\
Line ID &Fe K\nl
log(P)&-3.49\nl
Equivalent Width (eV)&115$^+_-$40\nl
$\chi^2$/PHA bins&718/595\\
\enddata
\end{deluxetable}

\begin{deluxetable}{lll} 
\tablenum{2}
\tablecolumns{3} 
\tablewidth{0pc} 
\tablecaption{Results of fitting to Chandra HETG}
\tablehead{
\colhead{Parameter} & \colhead{Power Law Only} & \colhead{Power Law + Gaussian} }
\startdata 
NH ($\times 10^{21}$ cm$^{-2}$) &1.6&1.6\\
photon index& 1.00$^{+0.15}_{-0.9}$&1.09$^{+0.15}_{-0.9}$\\
normalization( erg cm$^{-2}$ s$^{-1}$)&.0031$^{+.0007}_{-.0006}$&.0033$^{+.0007}_{-.0006}$\\
line 1&&\\
wavelength ($\AA$) &--&12.16\\
energy (keV) &--&1.02$^{+0.0015}_{-0.00025}$\\
normalization ($\times 10^{-5}$erg cm$^{-2}$ s$^{-1}$)&--&2.4$^{+2}_{-1.5}$\\
equivalent width (eV)&--&6.65\\ 
ID&--&Ne X L$\alpha$\\
log(P)&--&-19.2\\
line 2&&\\
wavelength ($\AA$) &--&10.60\\
energy (keV) &--&1.17$^+_-$0.01\\
normalization ($\times 10^{-5}$erg cm$^{-2}$ s$^{-1}$) &--&1.1$^{+1.2}_{-0.8}$\\
equivalent width (eV)&--&3.52\\ 
ID&--&Ne X L$\beta$\\
log(P)&--&-7.74\\
line 3&&\\
wavelength ($\AA$) &--&8.56\\
energy (keV) &--&1.45$^+_-$0.0025\\
normalization ($\times 10^{-5}$erg cm$^{-2}$ s$^{-1}$) &--&7.0$^{+5}_{-5.}$\\
equivalent width (eV)&--&0.185\\ 
ID&--&Mg XII L$\alpha$\\
log(P)&--&-0.06\\
line 4&&\\
wavelength ($\AA$) &--&6.18\\
energy (keV) &--&2.006$^{+0.0003}_{-0.004}$\\
normalization ($\times 10^{-5}$erg cm$^{-2}$ s$^{-1}$) &--&1.95$^{+0.1}_{-0.2}$\\
equivalent width (eV)&--&10.2\\ 
ID&--&Si XIV L$\alpha$\\
log(P)&--&-34.3\\
line 5&&\\
wavelength ($\AA$) &--&4.73\\
energy (keV) &--&2.62$^+_-$0.01\\
normalization ($\times 10^{-5}$erg cm$^{-2}$ s$^{-1}$) &--&1.03$^{+1.5}_{-1.}$\\
equivalent width (eV)&--&13.2\\ 
ID&--&S XVI L$\alpha$\\
log(P)&--&-4.09\\
\enddata 
\end{deluxetable} 

\clearpage

\begin{deluxetable}{lll} 
\tablenum{2}
\tablecolumns{3} 
\tablewidth{0pc} 
\tablecaption{Results of fitting to Chandra HETG, continued}
\tablehead{
\colhead{Parameter} & \colhead{Power Law Only} & \colhead{Power Law + Gaussian} }
\startdata 
line 6&&\\
wavelength ($\AA$) &--&1.938\\
energy (keV) &--&6.400$^+_-$0.01\\ 
normalization ($\times 10^{-5}$erg cm$^{-2}$ s$^{-1}$) &--&1.1$^+_- 2$ \\
equivalent width (eV)&--&26.2\\ 
ID&--&Fe neutral-like K$\alpha$\\
log(P)&--&-2.03\\
line 7&&\\
wavelength ($\AA$) &--&1.865\\
energy (keV) &--&6.650$^+_-$0.01\\
normalization ($\times 10^{-5}$erg cm$^{-2}$ s$^{-1}$) &--&1.4$^+_- 2$ \\
equivalent width (eV)&--&31.1\\ 
ID&--&Fe XXV 1-2\\
log(P)&--&-3.28\\
line 8&&\\
wavelength ($\AA$) &--&1.797\\
energy (keV) &--&6.900$^+_-$0.01\\
normalization ($\times 10^{-5}$erg cm$^{-2}$ s$^{-1}$) &--&$\leq 2$\\
equivalent width (eV)&--&50.1\\ 
ID&--&Fe XXVI L$\alpha$\\
log(P)&--&-0.2\\
C statistic/PHA bins&9390/8384&9009/8384\\
\enddata 
\end{deluxetable} 

\clearpage

\begin{deluxetable}{ll} 
\tablenum{3}
\tablecolumns{2} 
\tablewidth{0pc} 
\tablecaption{Results of fitting to Chandra HETG with xstar table + power law model}
\tablehead{ 
\colhead{Parameter} & \colhead{Value} }
\startdata 
NH ($\times 10^{21}$ cm$^{-2}$) &2.0\\
photon index& 1.03$^{+0.15}_{-0.9}$\\
normalization (erg cm$^{-2}$ s$^{-1}$)& .00286$^{+.0007}_{-.0006}$\\
xstar column density&10$^{21}$ cm$^{-2}$\\
log($\xi$)&4.5\\
xstar normalization&300.0$^{+300}_{-200}$\\
C statistic/PHA bins&8731/8384\\
\enddata 
\end{deluxetable} 

\clearpage

\begin{deluxetable}{lll} 
\tablenum{4}
\tablecolumns{3} 
\tablewidth{0pc} 
\tablecaption{Results of fitting to XMM EPIC MOS}
\tablehead{ 
\colhead{Parameter} & \colhead{Power Law only} & \colhead{power law + Gaussian} }
\startdata 
NH ($\times 10^{21}$ cm$^{-2}$) &1.735$^+_-$0.05&1.757$^+_-$0.05\\
photon index& 1.184$+_-$0.01& 1.193$+_-$0.01\\
normalization ($\times 10^{-3}$ erg cm$^{-2}$ s$^{-1}$& 6.7$^+_-$0.15& 6.78$^+_-$0.15\\
line 1&\\
wavelength ($\AA$) &--&1.85\\
energy (keV) &--&6.698$^+_-$0.1\\
normalization ($\times 10^{-5}$erg cm$^{-2}$ s$^{-1}$) &--&4.0$^+_-$1.0\\
equivalent width (eV) &--&1500\\
ID &--& Fe XXV 1 -- 2\\
log(P)&--&-7.5\\
line 2&\\
wavelength ($\AA$) &--&1.78\\
energy (keV) &--&6.978$^+_-$0.1\\
normalization ($\times 10^{-5}$erg cm$^{-2}$ s$^{-1}$) &--&1.2$^+_-$1.0\\
equivalent width (eV) &--&288\\
ID &--&Fe XXVI L$\alpha$\\
log(P)&--&-1.95\\
$\chi^2/\nu$&2077/1209&1969/1203\\
\enddata 
\end{deluxetable} 

\clearpage

\begin{deluxetable}{lll} 
\tablenum{5}
\tablecolumns{3} 
\tablewidth{0pc} 
\tablecaption{Results of fitting to XMM EPIC PN}
\tablehead{ 
\colhead{Parameter} & \colhead{Power Law Only} & \colhead{Power Law + Gaussian} }
\startdata 
NH ($\times 10^{21}$ cm$^{-2}$)& 6.6&11.2\\
photon index& 1.2&1.2\\
High Energy Cutoff (keV)&--&7\\
normalization ( erg cm$^{-2}$ s$^{-1}$)& .0112$^+_-$0.005& .0116$^+_-.0003$\\
line 1&&\\
wavelength ($\AA$) &--&1.94\\
energy (keV) &--&6.38$^+_-$0.1\\
width (keV)&--&0.29$^+_-$0.15\\
normalization ($\times 10^{-5}$erg cm$^{-2}$ s$^{-1}$) &--&1.4$^{+0.15}_{-0.05}$\\
equivalent width (eV) &--&14\\
ID &--&Fe I -- XVII K$\alpha$\\
log(P)&--&-99.\\
line 2&&\\
wavelength ($\AA$) &--&1.86\\
energy (keV) &--&6.68$^+_-$0.03 \\
width (keV)&--&$\leq$0.1 keV\\
normalization ($\times 10^{-5}$erg cm$^{-2}$ s$^{-1}$) &--&5.7$^+_-$0.2\\
equivalent width (eV) &--&63\\
ID &--&Fe XXV 1-2\\
log(P)&--&-5.59\\
\enddata 
\end{deluxetable} 

\clearpage

\begin{deluxetable}{lll} 
\tablenum{5}
\tablecolumns{3} 
\tablewidth{0pc} 
\tablecaption{Results of fitting to XMM EPIC PN, continued}
\tablehead{ 
\colhead{Parameter} & \colhead{Power Law Only} & \colhead{Power Law + Gaussian} }
\startdata 
line 3&&\\
wavelength ($\AA$) &--&1.78\\
energy (keV) &--&6.96$^{+0.02}_{-0.07}$\\
width (keV)&--&0.05$^+_-$0.04\\
normalization ($\times 10^{-5}$erg cm$^{-2}$ s$^{-1}$) &--&4.0$^+_-0.1$\\
equivalent width (eV) &--&49\\
ID &--& Fe XXVI L$\alpha$\\
log(P)&--&-99.\\
line 4&&\\
wavelength ($\AA$) &--&1.59\\
energy (keV) &--&7.8$^+_-0.1$\\
width (keV)&--&$\leq$ 0.5\\
normalization ($\times 10^{-5}$erg cm$^{-2}$ s$^{-1}$) &--&1.3$^+_-0.3$\\
equivalent width (eV) &--&23\\
ID &--&--\\
log(P)&--&-3.74\\
line 5&&\\
wavelength ($\AA$) &--&1.51\\
energy (keV) &--&8.2$^+_-0.1$\\
width (keV)&--&$\leq$ 0.3\\
normalization ($\times 10^{-5}$erg cm$^{-2}$ s$^{-1}$) &--&1.4$^+_-0.3$\\
equivalent width (eV) &--&25\\
ID &--&--\\
log(P)&--&-5.58\\
$\chi^2/\nu$&2074/1039&874/1028\\
\enddata 
\end{deluxetable} 

\clearpage

\begin{deluxetable}{lll} 
\tablenum{6}
\tablecolumns{3} 
\tablewidth{0pc} 
\tablecaption{Results of fitting to EPIC RGS}
\tablehead{ 
\colhead{Parameter} & \colhead{Power Law Only} &  \colhead{Power Law + Gaussians} }
\startdata 
NH ($\times 10^{21}$ cm$^{-2}$) &2.1&2.1\\
photon index& 1.2&1.2\\
normalization (erg cm$^{-2}$ s$^{-1}$)& .0085$^+_-0.0001$&  .0085$^+_-0.0001$\\
line 1&&\\
wavelength ($\AA$) &--&23.22\\
energy (keV) &--&0.534$^+_-$0.01\\
normalization ($\times 10^{-5}$erg cm$^{-2}$ s$^{-1}$) &--&5.3$^+_-$7\\
equivalent width (eV) &--&2.8\\
ID &--& --\\
log(P)&--&-1.06\\
line 2&&\\
wavelength ($\AA$) &--&22.92\\
energy (keV) &--&0.541$^+_-$0.01\\
normalization ($\times 10^{-5}$erg cm$^{-2}$ s$^{-1}$) &--&7.1$^+_-$8\\
equivalent width (eV) &--&3.5\\
ID &--& --\\
log(P)&--&-1.34\\
line 3&&\\
wavelength ($\AA$) &--&22.10\\
energy (keV) &--&0.561$^+_-$0.01\\
normalization ($\times 10^{-5}$erg cm$^{-2}$ s$^{-1}$) &--&4.5$^+_-$5\\
equivalent width (eV) &--&2.2\\
ID &--& O VII 1s$^2$($^1$S)-1s2s($^3$S)\\
log(P)&--&-1.15\\
line 4&&\\
wavelength ($\AA$) &--&21.77\\
energy (keV) &--&0.569$^+_-$0.01\\
normalization ($\times 10^{-5}$erg cm$^{-2}$ s$^{-1}$) &--&5.9$^+_-$12\\
equivalent width (eV) &--&3.3\\
ID &--&O VII 1s$^2$($^1$S)-1s2p($^3$P)\\
log(P)&--&-2.27\\
\enddata 
\end{deluxetable} 

\clearpage

\begin{deluxetable}{lll} 
\tablenum{6}
\tablecolumns{3} 
\tablewidth{0pc} 
\tablecaption{Results of fitting to EPIC RGS, continued}
\tablehead{ 
\colhead{Parameter} & \colhead{Power Law Only} &  \colhead{Power Law + Gaussians} }
\startdata 
line 5&&\\
wavelength ($\AA$) &--&19.0\\
energy (keV) &--&0.652$^+_-$0.01\\
normalization ($\times 10^{-5}$erg cm$^{-2}$ s$^{-1}$) &--&7.6$^+_-$2\\
equivalent width (eV) &--&5.4\\
ID &--&O VIII L$\alpha$\\
log(P)&--&-14.35\\
line 6&&\\
wavelength ($\AA$) &--&15.97\\
energy (keV) &--&0.775$^{+0.0015}_{-0.00025}$\\
normalization ($\times 10^{-5}$erg cm$^{-2}$ s$^{-1}$) &--&1.4$^+_-$2\\
equivalent width (eV) &--&1.2\\
ID &--& O VIII L$\beta$\\
log(P)&--&-0.84\\
line 7&&\\
wavelength ($\AA$) &--&12.13\\
energy (keV) &--&1.021$^{+0.0003}_{-0.004}$\\
normalization ($\times 10^{-5}$erg cm$^{-2}$ s$^{-1}$) &--&3.6$^{+2.3}_{-2.3}$\\
equivalent width (eV) &--&3.8\\
ID &--& Ne X L$\alpha$\\
log(P)&--&-2.17\\
line 8&&\\
wavelength ($\AA$) &--&10.65\\
energy (keV) &--&1.167$^{+0.0003}_{-0.004}$\\
normalization ($\times 10^{-5}$erg cm$^{-2}$ s$^{-1}$) &--&2.0$^{+1.5}_{-0.}$\\
equivalent width (eV) &--&2.9\\
ID &--&Ne X L$\beta$\\
log(P)&--&-0.60\\
line 9&&\\
wavelength ($\AA$) &--&8.40\\
energy (keV) &--&1.474$^{+0.0003}_{-0.004}$\\
normalization ($\times 10^{-5}$erg cm$^{-2}$ s$^{-1}$) &--&4.0$^{+2.7}_{-2.7}$\\
equivalent width (eV) &--&7.3\\
ID &--& Mg XII L$\alpha$\\
log(P)&--&-3.09\\
$\chi^2/\nu$&1522/1672&1393/1650\\
\enddata 
\end{deluxetable} 

\clearpage

\begin{deluxetable}{ll} 
\tablenum{7}
\tablecolumns{2} 
\tablewidth{0pc} 
\tablecaption{Results of fitting to XMM RGS with xtar table + power law model}
\tablehead{ 
\colhead{Parameter} & \colhead{Value} }
\startdata 
NH ($\times 10^{21}$ cm$^{-2}$) &2.2\\
photon index& 1.2\\
normalization (erg cm$^{-2}$ s$^{-1}$) & .0081$^{+.0007}_{-.0006}$\\
xstar column density&10$^{21}$ cm$^{-2}$\\
xstar density&10$^{11}$ cm$^{-2}$\\
Component 1&\\
log($\xi$)&1\\
xstar normalization&2.0$\times 10^{-4}$\\
Component 2&\\
log($\xi$)&4.5\\
xstar normalization&26.9\\
$\chi^2/\nu$&1413/1670\\
\enddata 
\end{deluxetable} 

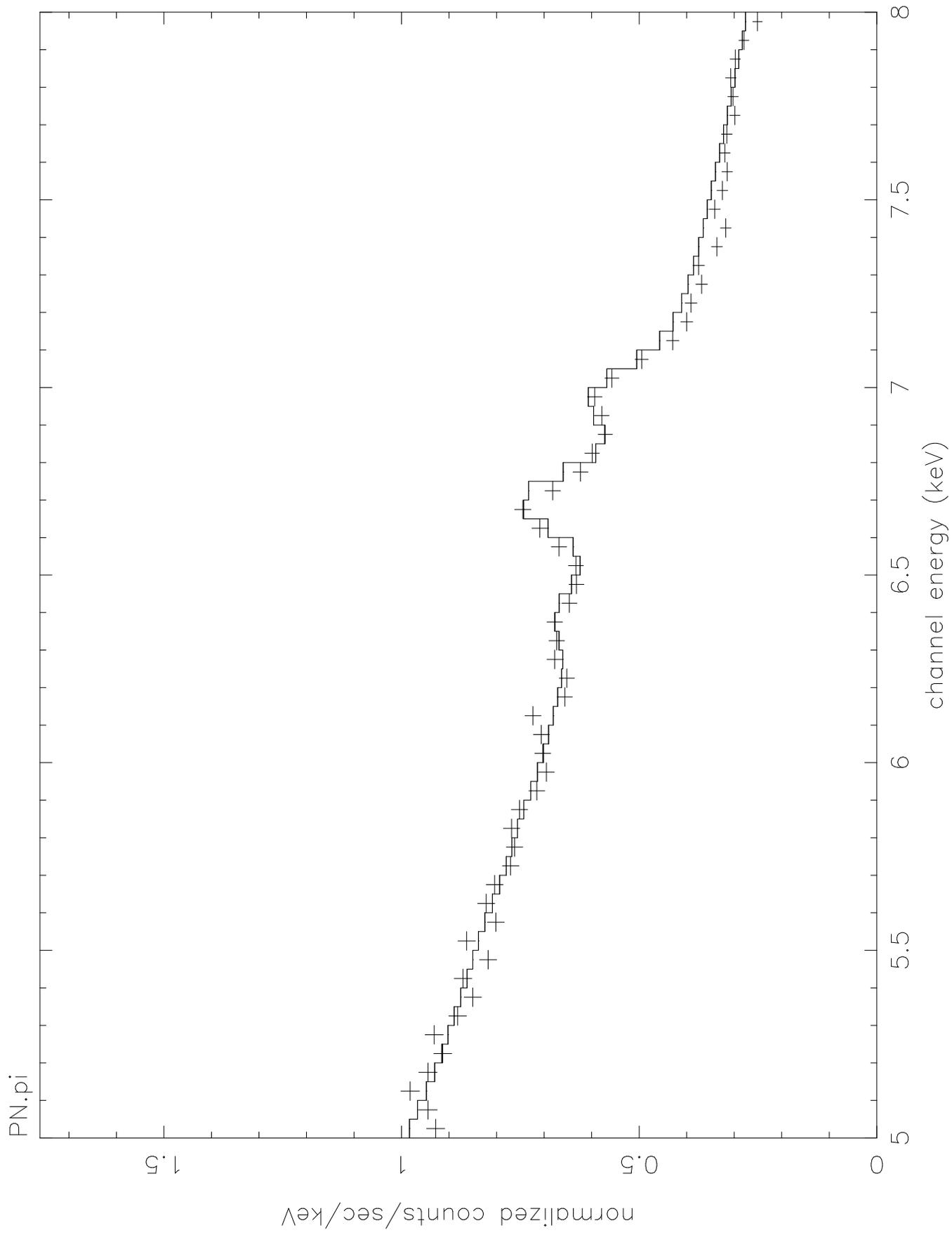
\end{document}